\documentstyle[prl,twocolumn,aps,floats,epsfig]{revtex}

\begin{document}

\draft
\twocolumn[\hsize\textwidth\columnwidth\hsize\csname
@twocolumnfalse\endcsname

\title{Nonlinear excitations: Solitons} 

\author{A. Mourachkine} 

\address{Nanoscience Centre and the Cavendish Laboratory, 
University of Cambridge, 11 J. J. Thomson Avenue, 
Cambridge CB3 0FF, UK} 

\maketitle

\begin{abstract} 

The main purpose of this chapter is to present a brief introduction to 
nonlinear excitations, and to underline the soliton concept. This chapter is 
Chapter 5 in the book {\it High-Temperature
Superconductivity: The Nonlinear Mechanism and Tunneling Measurements}  
(Kluwer Academic Publishers, Dordrecht, 2002), pages 101-142. 

\end{abstract}

\pacs{\hspace*{5mm} {\bf  True laws of Nature cannot be linear.} - Albert Einstein} 
]

\vspace*{2mm}

One may wonder what this Chapter is doing in a book describing the 
phenomenon of superconductivity. 
The main purpose of this Chapter is to present a brief introduction to 
nonlinear excitations, and to underline the soliton concept. Acquaintance 
with nonlinear excitations is necessary since the Cooper pairs in 
high-$T_{c}$ superconductors are pairs of soliton-like excitations. 
I am sure that, in 15 years or so, the presence of this Chapter 
in a book describing the phenomenon of superconductivity will be 
considered absolutely natural. 
There are excellent books devoted to the description of solitons and 
related phenomena; the reader who is interested knowing more on the 
soliton issues is referred to a few books (see Appendix). 

\section{Introduction} 

For a long time linear equations have been used for describing different
phenomena. For example, Newton's, Maxwell's and Schr\"odinger's
equations are linear, and they take into account only a linear response of
a system to an external disturbance. However, the majority of real systems 
are {\em nonlinear}. Most of the theoretical models are still relying on a
{\em linear} description, corrected as much as possible for 
nonlinearities which are treated as small perturbations. It is well known 
that such an approach can be absolutely wrong. The linear approach can 
sometimes miss completely some essential behaviors of the system. 

{\em Nonlinearity} has to do with {\it thresholds}, with multi-stability, 
with hysteresis, with phenomena which are changed {\em qualitatively} as 
the excitations are changed. 

In a linear system, the ultimate effect of the combined action of two 
different causes is merely the superposition of the effects of each cause 
taken individually. But in a nonlinear system adding two elementary actions 
to one another can induce dramatic new effects reflecting the onset of 
cooperativity between the constituent elements. 

To understand {\em nonlinearity}, one should first understand 
{\em linearity}. Consider linear waves. 
In general, a {\em wave} may be defined as a progression through matter of a 
state of motion. Characteristic properties of any linear wave are: (i) the 
shape and velocity of a linear wave are independent of its amplitude; (ii) the 
sum of two linear waves is also a linear wave; and (iii) small amplitude 
waves are linear. Figure 1a shows an example of a periodic linear wave. 
Large amplitude waves may become nonlinear. 
\begin{figure}[t]
\epsfxsize=0.7\columnwidth
\centerline{\epsffile{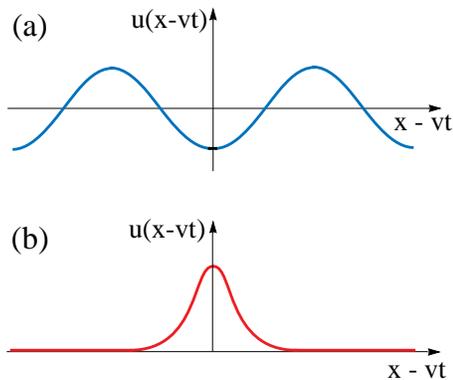}}
\caption{Sketch of (a) a periodic linear wave, and (b) a solitary wave.}
\end{figure} 

The fate of a wave travelling in a medium is determined by properties of the 
medium. {\em Nonlinearity} results in the distortion of the shape of large 
amplitude waves, for example, in turbulence. However, there is another source 
of distortion---the {\em dispersion} of a wave. More than 100 years ago the 
mathematical equations describing {\em solitary waves} were solved, 
at which point it was recognized that the solitary wave, shown in Fig. 1b, 
may exist due to a precise balance between the effects of nonlinearity and 
dispersion. Nonlinearity tends to make the hill steeper (see Fig. 1b), while 
dispersion flattens it. The solitary wave lives ``between'' these two 
dangerous, destructive ``forces.'' Thus, {\em the balance between nonlinearity 
and dispersion is responsible for the existence of the solitary waves}. 
As a consequence, the solitary waves are extremely robust.

Solitary waves or {\em solitons} cannot be described by using linear 
equations. Unlike ordinary waves which represent 
a spatial periodical repetition of elevations and hollows on a water surface, 
or condensations and rarefactions of a density, or deviations from a mean 
value of various physical quantities, solitons are single elevations, such 
as thickenings etc., which propagate as a unique entity with a given 
velocity. The transformation and motion of solitons are described by
nonlinear equations of mathematical physics.

The history of solitary waves or solitons is unique. The first {\em scientific} 
observation of the solitary wave was made by Russell in 1834 on the 
water surface. One of the first mathematical equations describing solitary 
waves was formulated in 1895. And only in 1965 were solitary waves 
fully understood! Moreover, many phenomena which were well known before 
1965 turned out to be solitons! Only after 1965 was it realized that solitary 
waves on the water surface, nerve pulse, vortices, tornados and many others 
belong to the same category: they are all solitons! That is not 
all, the most striking property of solitons is that they behave like particles! 

Mathematically, there is a difference between ``solitons'' and ``solitary 
waves.'' {\em Solitons} are localized solutions of integrable equations, 
while {\em solitary waves} are localized solutions of non-integrable 
equations. Another characteristic feature of {\em solitons} is that they 
are solitary waves that are not deformed after collision with other solitons. 
Thus the variety of {\em solitary waves} is much wider than the variety of 
the ``true'' solitons. Some solitary waves, for example, vortices and 
tornados are hard to consider as {\em waves}. For this reason, they are 
sometimes called {\em soliton-like excitations}. To avoid this bulky 
expression we shall often use the term {\em soliton} in all cases. This is 
not dangerous when we are talking about general 
properties of soliton-like excitations. 

\section{Russell's discovery} 

The first observation of the solitary wave or soliton was 
made by John Scott Russell near Edinburgh (Scotland) in August 1834. He was 
observing a boat moving on a shallow channel and noticed that, when the boat 
suddenly stopped, the wave that it was pushing at its prow ``rolled forward 
with great velocity, assuming the form of a large solitary elevation, a 
rounded, smooth and well defined heap of water which continued its course 
along the channel apparently without change of form or diminution of speed'' 
[1]. He followed the wave along the channel for more than a mile. The shape 
of the solitary wave observed by Russell is similar to that in Fig. 1b. 

Russell published the first report of this event in 1838. He called this 
solitary wave the {\em Wave of Translation}. A more detailed account of 
it and of successive experiments was published in his {\it Report on Waves} 
in 1844 [1]. 

Between 1834 and 1844, Russell performed numerous experiments in 
natural environments---on canals, rivers and lakes---as well as in his 
``laboratory,'' which was a specially designed small reservoir in his garden. 
In these studies, he found the following main properties of the solitary 
wave: 

\begin{itemize} 

\item 
An isolated solitary wave has a constant velocity and does not 
change its shape. 

\item 
The dependence of the velocity $\upsilon$ on the canal depth $h$ 
and the height of the wave, $u$, is given by the relation 

\[ \upsilon = \sqrt{g(h + u)}, \] 
where $g$ is the gravity acceleration. This formula is valid for $u < h$. 

\item 
A high enough solitary wave decays into two or more smaller 
solitary waves. The new ``born'' solitary waves have different heights and, 
as a consequence, their velocities are different.  

\item 
There exist only solitary elevations (humps); solitary cavities 
(depressions) are never met. 

\end{itemize} 

In his report, Russell made a remark that ``the great primary waves of 
translations cross each other without change of any kind in the same manner 
as the small oscillations produced on the surface of a pool by a falling stone''.  
Researchers were very much puzzled by this striking phenomenon which 
was understood only 130 years later.

Russell was not only an exceptionally observant scientist and excellent 
experimenter but a first class theorist. However, the attitude to solitons of 
some most prominent theoretical experts at that time was quite different, 
and his colleagues neither saw its significance nor shared his enthusiasm.
Moreover, there were even a few papers, including one written by Stokes, 
which concluded that the solitary wave cannot exist even in liquids with 
vanishing viscosity. 

Unfortunately, such a situation happens often in science. As noted in Chapter 
1, Abrikosov could not publish his famous paper during 5 years. The results 
described in the paper seemed to other physicists very strange. Even after 
1957, when it was published, the results were accepted 
only after experimental proof of several predicted effects. Everyone 
who is really active in science experienced in his (her) lifetime the same 
attitude from other scientists. ``I know how helpless an individual is against 
the spirit of his time,'' -- Ludwig Boltzmann. 

So, this was really a bad time for the solitary wave. Between 1844, when 
Russell's report was published, and 1965, fewer than two dozen papers 
relating to the solitary wave were published. 

\section{Korteweg--de Vries equation}

Fortunately, Russell lived long enough to see the solitary wave ``acquitted'' 
(he died in 1882). Boussinesq published a paper in 1871, in which he showed 
that Russell's solitary wave may exist and approximately calculated its 
shape and velocity. The final verdict to the existence of the solitary wave 
was  ``announced'' by Korteweg and de Vries in 1895. They reexamined all 
previous considerations and introduced a new equation for describing 
solitary waves that we now call the {\em Korteweg--de Vries} (KdV) 
equation [2]. 

Let $u(x,t)$ denote the height of the wave above the free surface at 
equilibrium, where $t$ is the time and $x$ is the coordinate along 
propagation of the solitary wave, the KdV equation is 

\begin{equation} 
\frac{\partial u}{\partial t} + \frac{3}{2h} u \frac{\partial u}{\partial x} 
+ \frac{h^{2}}{6} \frac{\partial ^{3} u}{\partial x^{3}} = 0, 
\end{equation}
where $h$ is the depth of the water in the canal. This equation is written 
in a frame moving at the speed of long-wave linear disturbances of the 
surface. The exact solution of the KdV equation describing the soliton is 

\begin{equation} 
u(x,t) = u_{0} \times sech ^{2} \left( \frac{x - \upsilon t}{\ell} \right), 
\end{equation} 
where $u_{0}$ is the initial height of the soliton; $\upsilon = \sqrt{gh} (1 + 
\frac{u_{0}}{2h})$, and 2$\ell$ is the ``width'' of the soliton and defined by 
the equation

\begin{equation} 
 S \equiv \frac{3}{4} \frac{u_{0} \ell ^{2}}{h^{3}} = 1. 
\end{equation} 

Equation (3) is the mathematical condition expressing the balance between the 
dispersion and nonlinearity effects in the soliton. Although the parameter 
$S$ was known for decades, its role in soliton theory became clear fairly 
recently. If $S$ is much larger than 1, nonlinearity effects will prevail (if 
the hump is smooth enough). They will strongly deform the hump and it will 
eventually break into several pieces that will probably give rise to several 
solitons. If $S <$ 1, dispersion prevails, and the hump will gradually diffuse. 
For $S \approx$ 1 the hump has a soliton-like shape and, if its velocity is 
close to the soliton velocity, it will slightly deform and gradually become 
the real soliton, described by the KdV equation.

The KdV equation played a crucial role in our times in resurrecting the 
soliton. For pure mathematicians, the history of solitons begins with the 
KdV equation. However, mathematicians did not fully realize the importance 
of the shallow water equation and neglected the KdV equation until the 
physicists returned it to life after 70 years of oblivion. In fact, Korteweg 
and de Vries themselves had no idea of the brilliant future awaiting their 
equation. 

\section{Numerical simulations}

With the appearance of computers, it became possible to simulate the 
behavior of nonlinear systems. Here we shall consider two first computer 
``experiments'': the first was performed by Fermi, Pasta and Ulam in 1955 
[3], and the second ten years later---in 1965 by Zabusky and Kruskal [4]. 
Historically, both simulations were very important for understanding 
the behavior of nonlinear systems described by the KdV equation. 

In 1955, Fermi, Pasta and Ulam (FPU) decided to investigate the behavior of 
a one-dimensional chain of 64 particles of mass $m$, bound by massless 
springs. They accounted for nonlinear forces by assuming that stretching the 
spring by $\Delta \ell$ generates the force 
$k \Delta \ell + \alpha ( \Delta \ell )^{2}$. 
The nonlinear correction to Hooke's law, $\alpha ( \Delta \ell )^{2}$, was 
assumed to be small as compared to the linear force, $k \Delta \ell$. Thus, 
the computer had to solve the following equations 
\begin{equation} 
m \ddot{u_{n}} = k(u_{n+1} - u_{n}) - k(u_{n} - u_{n-1}), 
\end{equation} 
with additional nonlinear terms in the right-hand side: 
\begin{equation} 
\alpha [(u_{n+1} - u_{n})^{2} - (u_{n} - u_{n-1})^{2}]. 
\end{equation} 
\begin{figure}[t]
\epsfxsize=0.98\columnwidth
\centerline{\epsffile{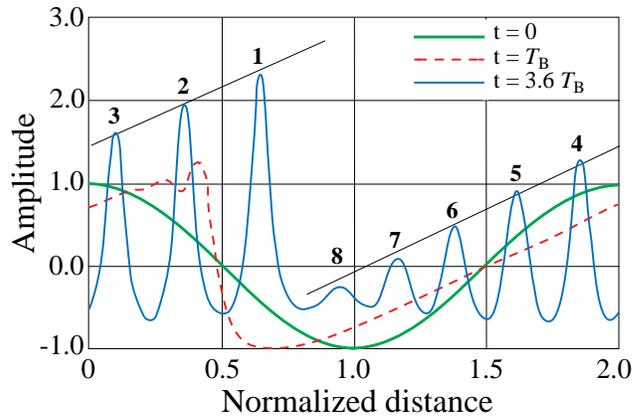}}
\caption{The temporal development of the wave form $u(x)$ [4].}
\end{figure}

In the harmonic case ($\alpha = 0$), the energy stored initially in a given 
mode stays in that mode and the system does not approach thermal 
equilibrium. For $\alpha \neq$ 0, Fermi, Pasta and Ulam started their 
simulation by exciting the lowest mode ($n$ = 1), i.e. by choosing a 
non-localized initial condition having the shape of the plane wave with 
wavelength equal to the size of the system. At the beginning of the 
simulation, they observed that the energy slowly goes into other modes, so 
the modes 2, 3, 4, ... became gradually excited. But, to their 
surprise, after about 157 periods of the fundamental mode, almost all the 
energy was back in the lowest mode!  After this recurrence time, the initial 
state was almost restored. This remarkable result, known as the FPU 
paradox, shows that introducing nonlinearity in a system does not guarantee 
an equipartition of energy. They also showed that nonlinear excitations can 
emerge not only from localized initial conditions but also from 
non-localized initial conditions or from thermal excitations.

When Kruskal and Zabusky heard of the results from the authors, they 
immediately started similar computer experiments: they considered 
movements of continuous nonlinear strings. After many attempts, they 
came to a striking conclusion: for small amplitudes, vibrations of the 
continuous FPU string are best described by the KdV equation! 

The KdV equation describes a variety of nonlinear waves, and is suitable 
for small amplitude waves in materials with weak dispersion. 

Zabusky and Kruskal solved the FPU paradox in 1965 because they 
plotted the displacements in {\em real space} instead of looking at the 
Fourier modes. Figure 2 shows the result of their simulations. They 
considered the evolution of a simple harmonic wave $u(0,x) = \cos( \pi x)$ 
shown by the green curve in Fig. 2. At some moment $T_{B}$ (= 1/$\pi$), 
a characteristic step is formed (see the dashed red curve), which later at time 
3.6$T_{B}$ gives rise to a sequence of solitons, presented by the solid curve. 
In Fig. 2, the solitons are enumerated in descending order, the first having 
the largest amplitude. The first soliton is also the fastest one, and it is 
running down other solitons and eventually colliding with them. Shortly, 
after each collision, the solitons reappear virtually unaffected in size and 
shape. Thus, Zabusky and Kruskal proved that the KdV solitons are not 
changed in collisions, like rigid bodies. This property of solitary waves was 
mentioned in Russell's report. Zabusky and Kruskal also found that, as two 
solitons pass through each other, they accelerate. As a consequence, their 
trajectories deviate from straight lines, meaning that the solitons have 
particle-like properties. We shall discuss this remarkable feature 
in the next Section. 

For the results of the simulation, shown in Fig. 2, the recurrence time 
$T_{R}$ was found to be $T_{R} = 30.4 T_{B}$. The FPU recurrence is simply 
a manifestation of the stability of the solitons. 
In fact, Zabusky and Kruskal studied waves in a plasma, which satisfy the 
KdV equation, and they coined the term {\it soliton}, 131 years after of its 
discovery.  

\section{Particle-like properties} 

Intuitively, it is natural to consider solitary waves are {\em waves}. However, 
what Zabusky and 
Kruskal discovered in their computer simulation is that solitons behave 
like particles. Historically, neither Russell nor other scientists who studied 
the solitary wave more than a century after him noticed its striking 
resemblance to a particle. As mentioned above, Russell was aware of the 
particle-like property of two colliding solitary waves---after collision both 
solitary waves preserve their shapes and velocities. However, Russell didn't 
notice that, when the higher wave overtake the lower one, it appears that the 
first simply goes through the second and moves on ahead. Russell obviously 
thought that this was true. In reality this is not the case. If Russell could 
have had a video camera, he would have seen that the process of the 
collision is somewhat different. 

When the larger solitary wave touches the smaller solitary wave, which 
move slower, the larger one slows down and diminishes while the smaller 
one accelerates and increases. When the smaller wave becomes as large as 
the original one, the waves detach and the ex-smaller one, which now is 
higher and faster, goes forward while the ex-bigger one is now moving 
behind at lower speed. Figure 3 schematically shows a collision of two 
solitary waves. In Fig. 3, two solitary waves are depicted at different 
periods of time. The observable result of the collision, shown in Fig. 3, is 
that the larger wave appears to have {\bf shifted} ahead of the 
position that it would 
have occupied in uniform motion (without the collision). While the smaller 
one appears correspondingly to have {\bf shifted} backwards. Thus, the 
waves do not penetrate each other, but rather collide like tennis balls. 
\begin{figure}[t]
\epsfxsize=0.85\columnwidth
\centerline{\epsffile{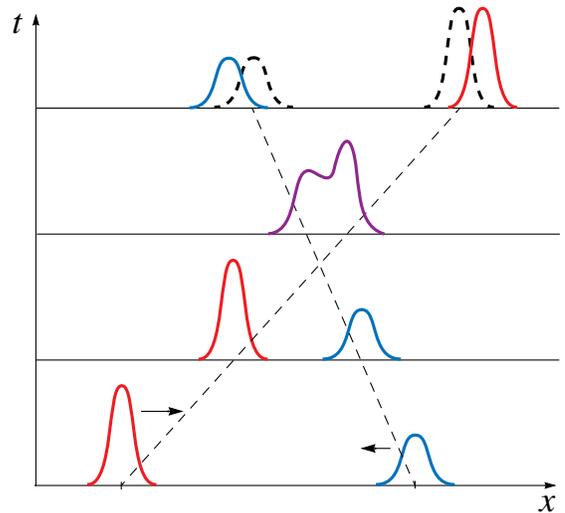}}
\caption{A schematic representation of a collision between two solitary 
waves.}
\end{figure} 

In order to understand this analogy, let us consider a collision of two tennis 
balls uniformly moving along the axis $x$. Suppose that the balls are 
identical, and they do not rotate. Let the velocities of the balls be 
$\upsilon _{1}$ and $\upsilon _{2}$. In the {\em center of mass system} of 
coordinates, which moves with the velocity $\upsilon = ( \upsilon _{1} - 
\upsilon _{2} )/2$, the balls have velocities $\upsilon$ and -$\upsilon$. 
Figure 4 shows two balls at different periods of time. Assume that the 
balls touch each other at time $t_{1}$ and detach at time $t_{2}$. Between 
$t_{1}$ and $t_{2}$, the balls suffer first squeezing and then expanding. What 
is interesting that, at the end of the collision, thus at $t_{2}$, the centers of 
the balls are slightly ahead (behind) of the positions $O'_{1}$ and $O'_{2}$ 
which correspond to the balls positions if there were no collision. This sort 
\begin{figure}[t]
\epsfxsize=0.9\columnwidth
\centerline{\epsffile{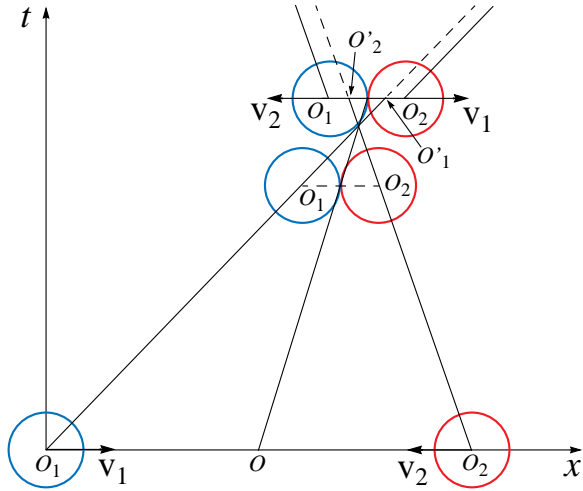}}
\caption{A schematic representation of a collision between two tennis 
balls [5].}
\end{figure} 
of shift occurs if the time of the collision ($t_{2} - t_{1}$) is smaller
than  the ``characteristic time'' 2$R/ \upsilon$, where $R$ is the radius
of the  balls. 

This experiment can be performed at home, if there is a 
fast-speed video camera. The question is why had no one noticed this 
striking particle-like property of the solitary wave? If Russell could not 
see this subtle effect because he didn't have proper equipment, it is more 
difficult to understand why experimenters using modern cinema equipment 
failed to observe the shift in 1952. The only reasonable explanation of this 
blindness of scientists is that everybody, including Russell, perceived the 
solitary wave as a {\em wave}. Even it was clear that this wave is 
very unusual, nobody could imagine regarding it as a particle. As soon as 
Zabusky and Kruskal found in their computer simulations that the solitary 
wave has much in common with particles, they cut off the word ``wave'' and 
gave the new name {\it soliton}, by analogy with electron, proton and other 
elementary particles. 
 
\section{Frenkel-Kontorova solitons} 

To finish historical introduction to the soliton, let us consider one more 
event which is important in the history of the soliton. 

All the examples of solitons, considered in the previous sections, are 
described by the KdV equation, and all these solitons belong to the 
same class of solitons: they are {\em nontopological}. The 
nontopological nature of these solitons can be easily understood because 
the system returns to its initial state after the passage of the wave. 
However, there are other types of solitons. The second group of solitons 
are so-called {\em topological} solitons, meaning that, after the passage 
of a topological soliton, the system is in a state which is different from 
its initial state. The topological stability can be explained by the analogy 
of the impossibility of untying a knot on an infinite  
rope without cutting it. 

Frenkel and Kontorova theoretically predicted in 1939 topological solitons 
[6]. In fact, they found a special sort of a defect, called a {\em dislocation}, 
which exist in the crystalline structure of solids. The dislocations are not 
immobile---they can move inside the crystal.

Frenkel and Kontorova studied the simple possible model of a crystal which 
is shown in Fig. 5. In this one-dimensional model, atoms (black balls in Fig. 
5) are distributed in a periodic sequence of hills and hollows which 
represent the substrate periodic potential. The balls rest at the bottom of 
each hollow because of the gravitational force. In Fig. 5, the springs which 
connect the balls represent the forces acting between atoms. It is evident 
that, in this model, a situation when one of the hollows is empty while all 
the balls are resting at the bottoms is impossible because of the springs. 
If one of the balls leaves a hollow (for example, $n$-th ball), the 
neighboring balls ($n$-1 and $n$+1) will be forced to follow. This
creates  an excitation which propagates further. The length of the 
\begin{figure}[t]
\epsfxsize=0.98\columnwidth
\centerline{\epsffile{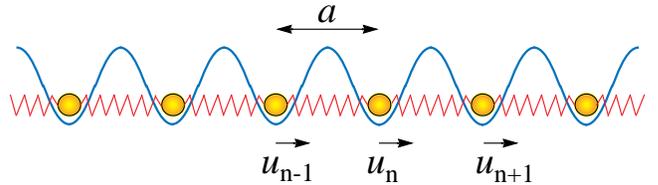}}
\caption{The Frenkel-Kontorova model of one-dimensional atomic chain: 
a chain of atoms with linear coupling (springs) interacting with a periodic 
nonlinear substrate potential.}
\end{figure}
excitation (or  dislocation) is much larger than the interatomic distance,
$a$. The long  dislocation is mobile because small shifts of each atom do
not require a  noticeable energy supply. So, the dislocations in an ideal
crystal freely  move without changing its shape. However, if the crystal
is imperfect, the  dislocations will be attracted or repulsed by the
defects.  It is not difficult to understand that two dislocations repel
each other, while  dislocations are attracted by antidislocations.  

The evolution of the dislocations and other topological solitons is described 
by the so-called {\em sine-Gordon} equation which we shall discuss in 
the next Section. The dislocation has the shape of a kink, shown in Fig. 6a. 
It has the {\em tanh}-like shape. The amplitude of the kink-soliton is 
independent of its velocity. As a consequence, topological solitons can 
be entirely static. The energy density of the kink-soliton is shown in Fig. 6b. 
In Fig. 6b, one can see that most of the energy of the kink, having the 
$sech^{2}$ shape, is localized in its core (in the middle of the kink). 
Consequently, {\em the soliton is a localized packet of energy}. The energy 
of the soliton falls exponentially away from its 
center. 
\begin{figure}[t]
\epsfxsize=0.7\columnwidth
\centerline{\epsffile{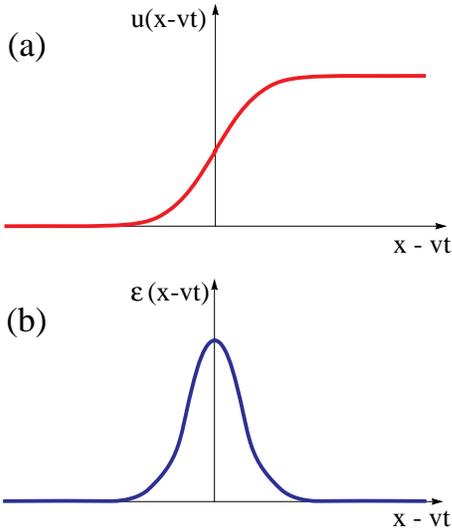}}
\caption{Schematic plots of (a) a kink-soliton solution, and (b) the energy 
density of the kink-soliton.}
\end{figure} 

\section{Topological solitons in a chain of pendulums} 

In order to understand better the nature of topological solitons, let us 
consider the propagation of a soliton in a chain of pendulums coupled by 
torsional springs. This mechanical transmission line is, probably, the 
simplest and one of the most efficient system for observing topological 
solitons and for studying their remarkable properties. 

Figure 7 shows 21 pendulums, each pendulum being elastically connected 
to its neighbors by springs. If the first pendulum is displaced by a small 
angle $\theta$, this disturbance propagates as a small amplitude linear 
wave from one pendulum to the next through the torsional coupling. As it 
moves along the chain, the small amplitude localized perturbation spreads 
over a larger and larger domain due to dispersive effects. The soliton is 
much more spectacular to observe. It is generated by moving the first 
pendulum by a full turn. This 2$\pi$ rotation propagates along the whole 
pendulum chain and, even, reflects at a free end to come back unchanged. 
\begin{figure}[t]
\epsfxsize=0.98\columnwidth
\centerline{\epsffile{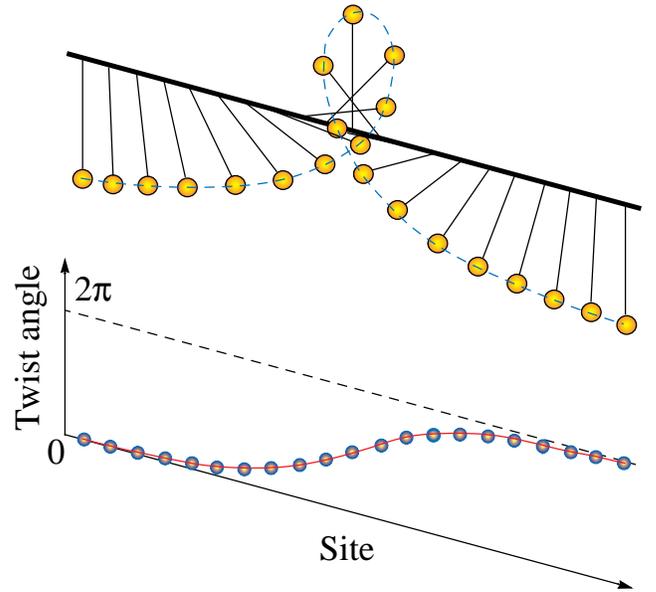}}
\caption{A soliton in a chain of pendulums coupled by a torsional spring. 
The soliton is a 2$\pi$ rotation. The lower figure shows the rotation angle 
of the pendulums as a function of their position along the chain.}
\end{figure}  

The pendulum chain involves only simple mechanics and it is easy to write 
its equations of motion. Its energy is the sum of the rotational kinetic 
energy, the elastic energy of the torsional string connecting two pendulums, 
and the gravitational potential energy. Denoting by $\theta _{n}$ the angle 
of deviation of pendulum $n$ from its vertical equilibrium position, by $a$ 
the distance between pendulums along the axis, by $m$ its mass and $L$ the 
distance between the rotation axis and its center of mass, the expression of 
the energy is then 

\begin{equation} 
H = \sum\limits_{n} \frac{1}{2} I \left( \frac{d \theta _{n}}{dt} \right) ^{2} 
+ \frac{1}{2} \beta ( \theta _{n} - \theta _{n-1})^{2} + 
mgL(1 - \cos \theta _{n}), 
\end{equation} 
where $I$ is the momentum of inertia of a pendulum around the axis, and 
$\beta$ is the torsional coupling constant of the springs. The equations of 
motion which can be deduced from the hamiltonian are 
\begin{equation} 
\frac{d^{2} \theta _{n}}{dt^{2}} - \frac{c_{0}^{2}}{a^{2}} (\theta _{n+1} + 
\theta _{n-1} - 2 \theta _{n}) + \omega ^{2}_{0} \sin \theta _{n} = 0, 
\end{equation} 
where $\omega _{0}^{2} = mgL/I$ and $c_{0}^{2} = \beta a^{2} /I$. 

This system of nonlinear differential equations, often 
called the {\em discrete sine-Gordon} equation, cannot be solved 
analytically. The common attitude in solving nonlinear equations is to 
{\em linearize} the equations. In our case, it can be done by replacing 
$\sin \theta _{n}$ by its small amplitude expression $\sin \theta_{n} 
\approx \theta _{n}$. Then the system of equations is easy to solve but 
{\em essential physics has been lost}. The linearized equations have no 
localized solutions and they have no chances to describe the soliton even 
approximately because $\theta _{n} = 2 \pi$ is not a small angle. Thus, 
{\em by linearizing a set of nonlinear equations to get an approximate 
solution is not always a good idea}. 

There is however another possibility to solve approximately this set of 
equations, while preserving their full nonlinearity, if the coupling between 
the pendulums is strong enough, i.e. $\beta \gg mgL$ (or $c_{0}^{2} /a^{2} \gg 
\omega _{0} ^{2}$). In this case, adjacent pendulums have similar motions 
and the discrete set of variables $\theta _{n} (t)$ can be replaced by a 
single function of two variables, $\theta (x,t)$ such that $\theta _{n} (t) = 
\theta (x = n, t)$. A Tylor expansion of $\theta (n + 1,t)$ and 
$\theta (n - 1,t)$ around $\theta (n,t)$ turns the discrete sine-Gordon 
equation into the partial differential equation 
\begin{equation} 
\frac{\partial ^{2} \theta (x,t)}{\partial t^{2}} - c_{0}^{2} 
\frac{\partial ^{2} \theta (x,t)}{\partial x^{2}} + 
\omega _{0} ^{2} \sin \theta = 0.
\end{equation} 
The equation is called the {\em sine-Gordon} equation, and it has been 
extensively studied in soliton theory because it has exceptional 
mathematical properties. In particular, it has a soliton solution 
\begin{equation} 
u(x - \upsilon t) = 4 \arctan \left[ \exp \left( \pm 
\frac{\omega _{0}}{c_{0}} \frac{x - \upsilon t}{\sqrt{1 - 
\upsilon ^{2}/c^{2}_{0}}} \right) \right],  
\end{equation} 
which is plotted in Figs. 7 and 8. The ($\pm$) signs correspond to 
localized soliton solutions which travel with the opposite screw senses: 
they are respectively called a {\em kink} soliton and an {\em antikink} 
soliton. They are shown in Fig. 8: the pendulums rotate from 0 to 
2$\pi$ for the kink and from 0 to -2$\pi$ for the antikink. 
\begin{figure}[t]
\epsfxsize=0.85\columnwidth
\centerline{\epsffile{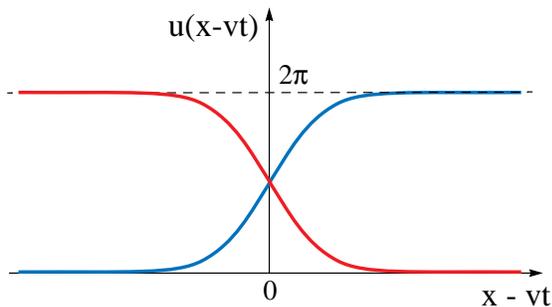}}
\caption{A schematic plot of a kink-soliton and antikink-soliton solutions 
for the chain of pendulums shown in Fig. 7.}
\end{figure} 

As the KdV equation, the sine-Gordon equation also contains dispersion and 
nonlinearity. However, in the sine-Gordon equation, both dispersion and 
nonlinearity appear in the same term 
$\omega _{0}^{2} \sin \theta$, where $\omega _{0}$ is a 
characteristic frequency of a system. 

The topological nature of solitons in the chain of pendulums can be 
easily demonstrated by plotting the gravity potential acting on the 
pendulums versus $\theta$ and the position $x$ of the pendulum. As one 
can see in Fig. 9, each kink joints two successive equilibrium states 
(potential minima). Thus, the kink-soliton can be considered as a 
{\em domain wall} between two degenerate energy minima. The topological 
soliton is an excitation which interpolates between these minima. 
\begin{figure}[t]
\epsfxsize=0.8\columnwidth
\centerline{\epsffile{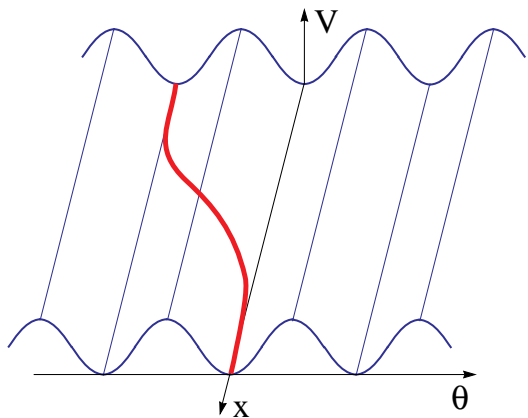}}
\caption{Sketch of the sinusoidal potential of the 
pendulum chain. The solid curve shows the trajectory of the kink-soliton 
which can be considered as a domain wall between two degenerate energy 
minima.}
\end{figure} 

In the expression of the soliton solution, one can see that a topological 
soliton cannot travel faster than $c_{0}$ which represents the velocity of 
linear waves (in solids, the longitudinal sound velocity). As the soliton 
velocity $\upsilon$ approaches $c_{0}$, the soliton remains constant but its 
width gets narrower owing to the Lorentz contraction of its profile, given by 
$d \sqrt{1 - \upsilon ^{2}/ c_{0}^{2}}$, where $d = c_{0} / \omega _{0}$ is 
a {\em discreteness} parameter. Thus, the topological solitons behave like 
relativistic particles. 

From the condition $c_{0}^{2} /a^{2} \gg \omega _{0} ^{2}$ (see above), 
the discreteness parameter $d = c_{0} / \omega _{0}$ is much larger than 
the distance between pendulums,  $d \gg a$. 
If $d \simeq a$, the angle of rotation varies abruptly from one pendulum to 
the next, and the continuum (long-wavelength) approximation cannot be used. 

Comparing topological and nontopological solitons, it is worth remarking 
that the amplitude of the kink is independent of its velocity and, when the 
velocity $\upsilon$ = 0, the soliton solution reduces to 
\begin{equation} 
u(x) = 4 \arctan [ \exp ( \pm x/d)].   
\end{equation} 
Thus, by contrast to nontopological solitons, {\em the kink soliton may be 
entirely static}, losing its wave character. 
The second feature of topological solitons (kinks) is that they have 
antisolitons (antikinks) which are analogous to antiparticles. In contrast, 
for nontopological solitons, there are no antisolitons.  

By using such a simple mechanical device shown in Fig. 7, it is easy to 
check experimentally the exceptional properties of the topological solitons. 
Launching a soliton and keeping on agitating the first pendulum, it is possible 
to test the ability of the soliton to propagate over a sea of linear waves. The 
particle-like properties appear clearly if one static soliton is created in the 
middle of the chain and then a second one is sent. The collision looks to the 
observer exactly similar to a shock between elastic marbles. Thus, the 
pendulum chain provides an experimental device which convincingly 
demonstrate the unique properties of the soliton. 

In addition to the kink and antikink solutions, 
the sine-Gordon equation also has solutions in the form of oscillating 
pulses called a {\em breather} or {\em bion} (meaning a ``living particle''). 
Breathers can be considered as a soliton-antisoliton bound state. An 
example of a breather is shown in Fig. 10. As other topological solitons, 
breathers can move with a constant velocity or be entirely static. 
Theoretically, breathers interact with other breathers and other solitons in 
the same manner as all topological solitons do. However, in reality, 
breather-type solitons can be easily destroyed by almost any type of 
perturbation. 
\begin{figure}[t]
\epsfxsize=0.8\columnwidth
\centerline{\epsffile{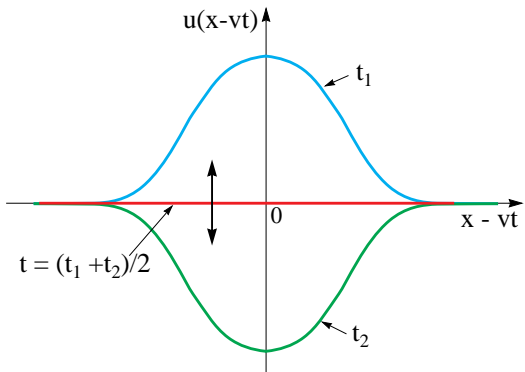}}
\caption{Sketch of a sine-Gordon breather or a bion at different times $t$.}
\end{figure} 

Stationary breathers are pulsating objects. If, in the chain of pendulums 
shown in Fig. 7, one launches a kink and antikink (with the opposite screw 
sense) with sufficiently low velocity, one can observe a bounded pair that 
is a breather solution. Nevertheless, owing to dissipation effects present on 
the real line, only a few breathing oscillations can be observed. The 
oscillations then decrease with time and energy is radiated onto the line. 

\section{Different categories of solitons} 

There are a few ways to classify solitons. For example, as discussed above, 
there are {\em topological} and {\em nontopological} solitons. Independently 
of the topological nature of solitons, all solitons can be divided into 
two groups by taking into account their profiles: {\em permanent} and {\em 
time-dependent}. For example, kink solitons have a permanent profile (in 
ideal systems), while all breathers have an internal dynamics, even, if they 
are static. So, their shape oscillates in time. The third way to classify the 
solitons is in accordance with nonlinear equations which describe their 
evolution. Here we discuss common properties of solitons on the basis of the 
third classification, i.e. in accordance with nonlinear equations which 
describe the soliton solutions. 

Up to now we have considered two nonlinear equations which are used to 
describe soliton solutions: the KdV equation and the sine-Gordon equation. 
There is the third equation which exhibits true solitons---it is called the 
{\em nonlinear Schr\"odinger} (NLS) equation. We now 
summarize 
soliton  properties on the basis of these three equations, namely, 
the Korteweg-de Vries equation: 
\begin{equation} 
u_{t} = 6uu_{x} - u_{xxx};
\end{equation} 
the sine-Gordon equation: 
\begin{equation} 
u_{tt} = u_{xx} - \sin u, 
\end{equation} 
and the nonlinear Schr\"odinger equation: 
\begin{equation} 
iu_{t} = - u_{xx} \pm |u|^{2}u, 
\end{equation} 
where $u_{z}$ means $\frac{\partial u}{\partial z}$. For simplicity, the 
equations are written for the dimensionless function $u$ depending on the 
dimensionless time and space variables. 

There are {\em many} other nonlinear equations (i.e. the Boussinesq 
equation) which can be used for evaluating solitary waves, 
however, these three equations are particularly important for physical 
applications. They exhibit the most famous solitons: 
the KdV (pulse) solitons, the sine-Gordon (topological) solitons 
and the {\em envelope} (or NLS) solitons. All the solitons are 
one-dimensional (or quasi-one-dimensional). Figure 11 schematically 
shows these three types of solitons. Let us summarize common features 
and individual differences of the three most important solitons. 
\begin{figure}[t]
\epsfxsize=0.95\columnwidth
\centerline{\epsffile{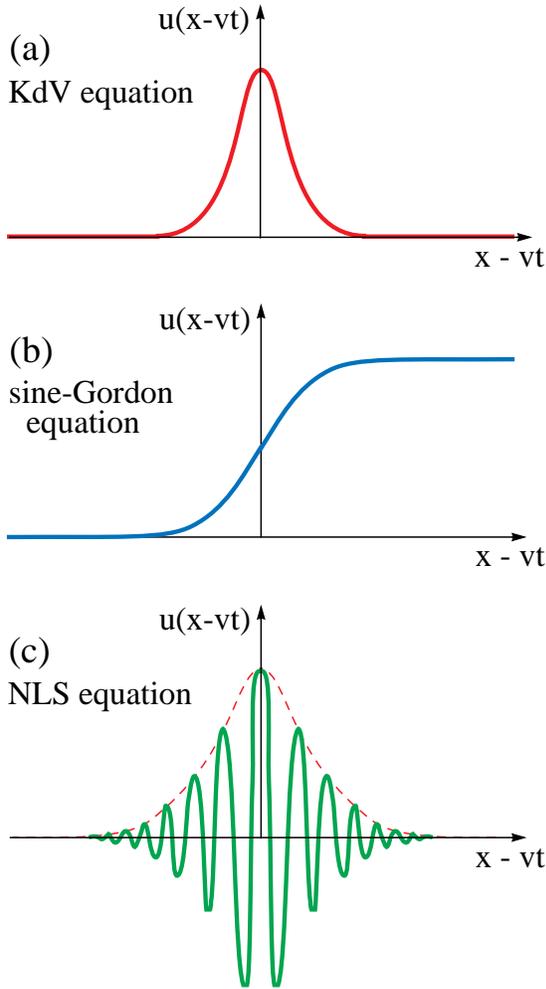}}
\caption{Schematic plots of the soliton solutions of: (a) the Korteweg--de 
Vries equation; (b) the sine-Gordon equation, and (c) the nonlinear 
Schr\"odinger equation.}
\end{figure} 

\subsection{The KdV solitons} 

The exact solution of the KdV equation is given by Eq. (2). 
The basic properties of the KdV soliton, shown in Fig. 11a, can be 
summarized as follows [7]: 

\begin{itemize} 

\item[i.] 
Its amplitude increases with its velocity (and vice versa). Thus, they 
cannot exist at rest. 

\item[ii.] 
Its width is inversely proportional to the square root of its velocity. 

\item[iii.]
It is a unidirectional wave pulse, i.e. its velocity cannot be negative 
for solutions of the KdV equation. 

\item[iv.]
The sign of the soliton solution depends on the sign of the nonlinear 
coefficient in the KdV equation. 

\end{itemize} 

The conservation laws for the KdV solitons represent the conservation of 
mass 
\begin{equation} 
M = \frac{1}{2} \int udx; 
\end{equation}  
momentum 
\begin{equation} 
P = - \frac{1}{2} \int u^{2}dx; 
\end{equation} 
energy 
\begin{equation} 
E = \frac{1}{2} \int (2u^{2} + u_{x}^{2})dx, 
\end{equation}  
and center of mass 
\begin{equation}  
Mx_{s}(t) = \frac{1}{6} \int xudx = tP + const. 
\end{equation}  
 
The KdV solitons are nontopological, and they exist in physical systems with 
weakly nonlinear and with weakly dispersive waves. When a wave impulse 
breaks up into several KdV solitons, they all move in the same direction 
(see, for example, Fig. 2). The collision of two KdV solitons is 
schematically shown in Fig. 3. Under certain conditions, the KdV solitons 
may be regarded as particles, obeying the standard laws of Newton's 
mechanics. In the presence of dissipative effects (friction), the KdV solitons 
gradually decelerate and become smaller and longer, thus, they are ``mortal.''  

\subsection{The topological solitons} 

The exact solution of the sine-Gordon equation is given by Eq. (9).
The basic properties of a topological (kink) soliton shown in Fig. 11b can be 
summarized as follows [7]: 

\begin{itemize} 

\item[i.] 
Its amplitude is independent of its velocity---it is constant and remains 
the same for zero velocity, thus the kink may be static. 

\item[ii.]
Its width gets narrower as its velocity increases, owing to Lorentz 
contraction. 

\item[iii.]
It has the properties of a relativistic particle. 

\item[iv.]
The topological kink which has a different screw sense is called an 
{\em antikink}. 

\end{itemize} 

For the chain of pendulums shown in Fig. 7, the energy of the topological 
(kink) soliton, $E_{K}$, is determined by 
\begin{equation} 
E_{K} = \frac{m_{0} c_{0}^{2}}{\sqrt{1 - \frac{\upsilon ^{2}}{c_{0}^{2}}}},  
\end{equation} 
where $c_{0}$ is the velocity of {\em linear} waves, and the soliton mass 
$m_{0}$ is given by 
\begin{equation} 
m_{0} = 8 \frac{I}{a} \frac{\omega _{0}}{c_{0}} = 8 \frac{I}{a} \frac{1}{d}.    
\end{equation} 

One can also introduce the relativistic momentum 
\begin{equation} 
p_{K} = \frac{m_{0} \upsilon}{\sqrt{1 - \frac{\upsilon ^{2}}{c_{0}^{2}}}}. 
\end{equation} 
At rest ($\upsilon$ = 0) one has $p$ = 0, and the static soliton energy is 
\begin{equation} 
E_{0,K} = m_{0}c_{0}^{2}. 
\end{equation} 

Topological solitons are extremely stable. Under the influence of 
friction, these solitons only slow down and eventually stop and, at rest, 
they can live ``eternally.'' In an infinite system, the topological soliton 
can only be destroyed by moving a semi-infinite segment of the system 
above a potential maximum. This would require an infinite energy. 
However, the topological soliton can be annihilated in a collision between 
a soliton and an anti-soliton. In an integrable system having exact soliton 
solutions, solitons and anti-solitons simply pass through each other 
with a phase shift, as all solitons do, but in a real system like 
the pendulum chain which has some dissipation of energy, the 
soliton-antisoliton equation may destroy the nonlinear excitations. 
Figure 12 schematically shows a collision of a kink and an antikink in an 
integrable system which has soliton solutions. 
In integrable systems, the soliton-breather and breather-breather collisions 
are similar to the kink-antikink collision shown in Fig. 12.  
\begin{figure}[t]
\epsfxsize=0.9\columnwidth
\hspace*{2mm}
\epsffile{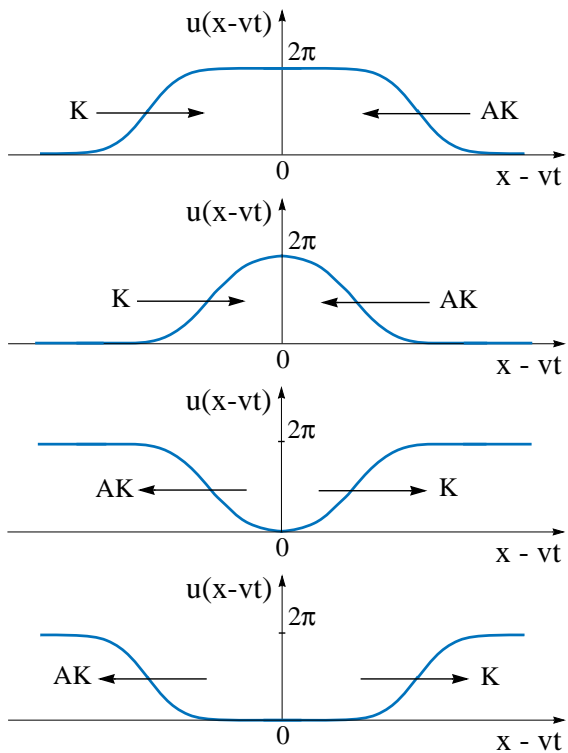} 
\caption{Sketch of a collision between a kink (K) and an 
antikink (AK). The phase shift after the collision is not indicated.}
\end{figure} 

The sine-Gordon equation has almost become ubiquitous in the theory of 
condensed matter, since it is the simplest nonlinear wave equation in a 
periodic medium. 

\subsection{The envelope solitons} 

The NLS equation is called the {\em nonlinear} Schr\"odinger
equation  because it is formally similar to the Schr\"odinger
equation of quantum  mechanics 
\begin{equation} 
\left[ i \hbar \frac{\partial}{\partial t} + \frac{\hbar ^{2}}{2m} 
\frac{\partial ^{2}}{\partial x^{2}} - U \right] \Psi (x,t) = 0, 
\end{equation} 
where $U$ is the potential, and $\Psi (x,t)$ is the wave function. 

The NLS equation describes self-focusing phenomena in nonlinear optics, 
one-dimensional self-modulation of monochromatic waves, in nonlinear 
plasma etc. In the NLS equation, the potential $U$ is replaced by $|u|^{2}$ 
which brings into the system self-interaction. The second term of the NLS 
equation is responsible for the dispersion, 
and the third one for the nonlinearity. 
A solution of the NLS equation is schematically shown in Fig. 11c. The 
shape of the enveloping curve (the dashed line in Fig. 11c) is given by 
\begin{equation} 
u(x,t) = u_{0} \times sech ((x- \upsilon t)/ \ell), 
\end{equation} 
where 2$\ell$ determines the width of the soliton. Its amplitude $u_{0}$ 
depends on $\ell$, but the velocity $\upsilon$ is {\it independent} of the 
amplitude, distinct from the KdV soliton. The shapes of the envelope and 
KdV solitons are also different: the KdV soliton has a $sech^{2}$ shape. 
Thus, the envelope soliton has a slightly wider shape. However, other 
properties of the envelope solitons are similar to the KdV solitons, thus, 
they are ``mortal'' and can be regarded as particles. The interaction between 
two envelope solitons is similar to the interactions between two KdV 
solitons (or two topological solitons), as shown in Fig. 3. 

In the envelope soliton, the stable 
groups have normally from 14 to 20 humps under the envelope, the central 
one being the highest one. The groups with more humps are unstable and 
break up into smaller ones. The waves inside the envelope move with a 
velocity that differs from the velocity of the soliton, thus, the envelope 
soliton has an {\em internal} dynamics. The relative motion of the envelope 
and carrier wave is responsible for the internal dynamics of the NLS soliton. 

The NLS equation is inseparable part of nonlinear optics where 
the envelope solitons are usually called {\em dark} and {\em bright} solitons, 
and became quasi-three-dimensional. We shall briefly discuss the optical 
solitons below.

\subsection{Solitons in real systems} 

As a final note to the presentation of the three types of solitons, it is 
necessary to remark that real systems do not carry exact soliton solutions 
in the strict mathematical sense (which implies an infinite life-time and an
infinity of conservation laws) but {\em quasi}-solitons which have most of 
the features of true solitons. In particular, although they do not have an 
infinite life-time, quasi-solitons are generally so long-lived that their 
effect on the properties of the system are almost the same as those of 
true solitons. This is why physicists often use the word soliton in a relaxed 
way which does not agree with mathematical rigor. 

In the following sections, we consider a few examples of solitons in real 
systems, which are useful for the understanding of the mechanism of 
high-$T_{c}$ superconductivity. 

\section{Solitons in the superconducting state}

Solitons are literally everywhere. The superconducting state is not an 
exception: vortices and {\em fluxons} are topological solitons. Fluxons 
are quanta of magnetic flux, which can be studied in {\em long Josephson 
junctions}. A long Josephson junction is analogous to the chain of atoms 
studied by Frenkel and Kontorova. The sine-Gordon equation provides an 
accurate description of vortices and fluxons. 

In a type-II superconductor, if the magnitude of an applied magnetic field 
is larger than the lower critical field, $B_{c1}$, the magnetic field begins to 
penetrates the superconductor in microscopic vortices which form a regular 
lattice. Vortices in a superconductor are similar to vortices in the ideal 
liquid, but there is a dramatic difference---they are {\em quantized}. 
Inside the vortex tube, there exists a magnetic field. 
The magnetic flux supported by the vortex is a multiple of  the 
{\em quantum magnetic flux} $\Phi _{0} \equiv hc/2e$, where $c$ is the 
speed of light; $h$ is the Planck constant, and $e$ is the electron charge. 
The vortices in type-II superconductors are a pure 
and beautiful physical realization of the sine-Gordon (topological) solitons. 
Their extreme stability can be easily understood in terms of the 
Ginzburg-Landau theory of superconductivity which contains a system of 
nonlinear coupled differential equations for the vector potential and the 
wave function of the superconducting condensate. Thus, topological 
solitons in the form of vortices were found in the superconducting state 
20 years earlier than the superconducting state itself was understood.  

Let us now consider fluxons in a long Josephson junction. An example of a 
{\em long} superconductor-insulator-superconductor (SIS) Josephson 
junction is schematically shown in Fig. 13. The insulator in the junction is thin
enough,  say 10--20 \AA, to ensure the overlap of the wave 
functions. Due to the  Meissner effect, the magnetic field may exist only in the
insulating layer  and in adjacent thin layers of the superconductor. 
\begin{figure}[t]
\epsfxsize=0.95\columnwidth
\centerline{\epsffile{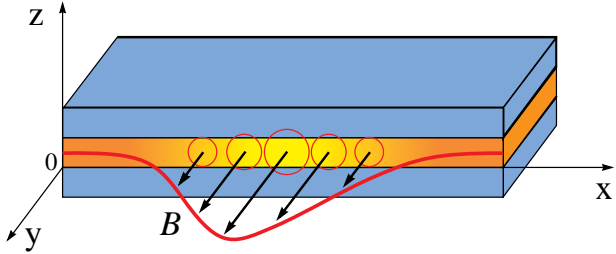}}
\caption{Sketch of a soliton in a long Josephson junction.}
\end{figure} 

As discussed in Chapter 2 in [14],  
the magnitude of the zero-voltage current resulting from the tunneling 
of Cooper pairs, known as the {\em dc} Josephson effect, depends on the 
phase difference between two superconductors as 
\begin{equation}
I = I_{c} \sin \varphi, 
\end{equation} 
where $\varphi = \varphi _{2} - \varphi _{1}$ is the phase difference, and 
$I_{c}$ is the critical Josephson current.

The oscillating current of Cooper 
pairs that flows when a steady voltage $V$ is maintained across a tunnel 
barrier is known as the {\em ac} Josephson effect. The phase difference 
between the two superconductors in the junction is given by 
\begin{equation}
\frac{d \varphi}{dt} = \frac{2e}{\hbar} V. 
\end{equation}  
Taking into account the inductance and the capacitance of the junction, one 
can easily get a sine-Gordon equation for $\varphi (x,t)$ forced by a 
right-hand side term which is associated with the applied bias. For a long 
Josephson junction, the nonlinear equation exactly coincides with Eq. (8). 
In the long Josephson junction, the 
sine-Gordon solitons describe quanta of magnetic flux, expelled from the 
superconductors, that travels back and forth along the junction. Their 
presence, and the validity of the soliton description, can be easily checked 
by the microwave emission which is associated with their reflection at the 
ends of the junction. 

For the Josephson junction, the ratio 
$\lambda_{J} = c_{0}/ \omega _{0}$ [see Eq. (8)] gives a measure of 
the typical distance over which the phase (or magnetic flux) changes, and is 
called the {\em Josephson penetration length}. This quantity allows one to 
define precisely a {\em small} and a {\em long} junction. A junction is said 
to be long if its geometric dimensions are large compared with 
$\lambda _{J}$. Otherwise, the junction is small. The Josephson length is 
usually much larger than the London penetration 
depth $\lambda _{L}$. 
 
The moving fluxon has a kink shape, which is accompanied by a 
{\em negative} voltage pulse and a {\em negative} current 
pulse, corresponding to the space and time derivatives of the fluxon solution. 
In Fig. 13, the arrows in the $y$ direction represent the magnetic field, and 
the circles show the Josephson currents producing the magnetic field. 

A very useful feature of the Josephson solitons 
is that they are not difficult to operate by applying bias and current to the 
junction. By using artificially prepared inhomogeneities one can make bound 
state of solitons. In this way one can store, transform and 
transmit information. In other words, long Josephson junctions can be used 
in computers. One of the most useful properties of such devices would be 
very high performance speed. Indeed, the characteristic time may be as 
small as 10$^{-10}$ sec, while the size of the soliton may be less than 0.1 
mm. The main problem for using the Josephson junctions in electronic 
devices is the cost of cooling refrigerators. However, I am absolutely sure 
that they will be commercially used in electronics if room-temperature 
superconductors become available. 

\section{Topological solitons in polyacetylene} 

Let us consider topological solitons and {\em polarons} in solids. 
One-dimensional polarons 
are breathers, consequently, they can be considered as a soliton-antisoliton 
(kink-antikink) bound state. The term ``polaron'' should not be confused with 
the Holtstein (small) polaron which is three-dimensional, and represents a 
self-trapped state in solids (see Section XII).

Polyacetylene, (CH)$_{x}$, is the simplest linear conjugated polymer. 
Polyacetylene exists in two isomerizations: {\em trans} and {\em cis}. Let 
us consider the {\em trans}\,-polyacetylene. The structure of 
{\em trans}\,-polyacetylene is schematically shown in Fig. 14. The 
important property of {\em trans}\,-polyacetylene is the double degeneracy 
of its ground state: the energy for the two possible patterns of alternating 
short (double) and long (single) bonds. The double-well potential is 
\begin{figure}[t]
\epsfxsize=0.8\columnwidth
\centerline{\epsffile{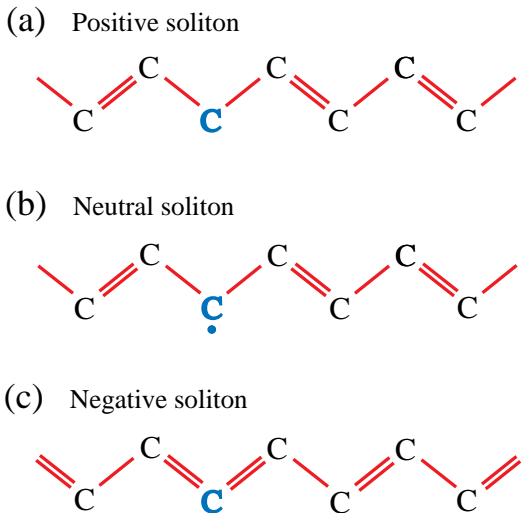}}
\caption{Formula of polyacetylene showing the possibility to have a 
topological soliton: (a) positive; (b) neutral, and (c) negative.}
\end{figure}
schematically shown in Fig. 15. In double-well potential, a kink cannot be 
followed by another kink like, for example, in a potential with infinite 
number of equilibrium states, shown in Fig. 9: a kink can be followed only 
by an antikink, which connect the two potential wells, as shown in Fig. 15. 
In {\em cis}\,-polyacetylene, the two ground-state configurations have 
different energies. Therefore, in {\em cis}\,-polyacetylene, the two-well 
potential is asymmetrical. Further we consider only the 
{\em trans}\,-polyacetylene configuration, sometimes, dropping the prefix 
``trans''.  

Undoped polyacetylene which has a half-filled band is an insulator with a 
charge gap of 1.5 eV. The gap is partially attributed to the so-called Peierls 
instability of the one-dimensional electron gas. Upon doping, polyacetylene 
becomes highly conducting. The significant overlap between the orbitals 
of the neighboring carbon atoms results in a relatively broad band (called 
\begin{figure}[t]
\epsfxsize=0.77\columnwidth
\centerline{\epsffile{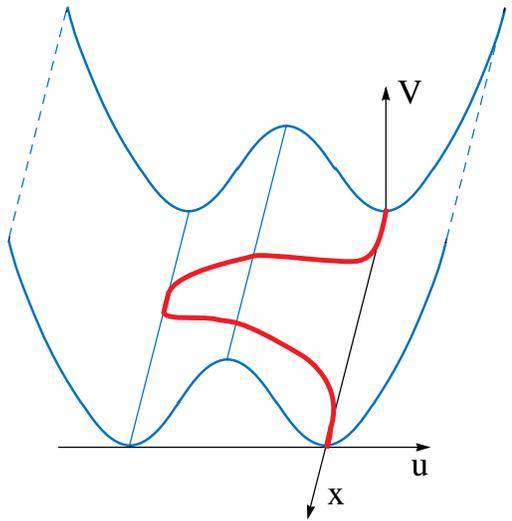}}
\caption{Sketch of the double-well potential for 
{\em trans}\,-polyacetylene. The solid curve shows the trajectory 
of the kink-antikink (breather) solution.}
\end{figure} 
$\pi$-band) with a width of 10 eV. The polyacetylene chains are only 
weakly coupled to each other: the distance between the nearest carbon 
atoms belonging to different chains is about 4.2 \AA, which is three times 
larger than the distance between the nearest carbon atoms in the chains. As 
a consequence, the interchain hopping amplitude is about 30 times smaller 
than the hopping amplitude along the chain, which makes the intrinsic 
properties of polyacetylene quasi-one-dimensional.
  
As a consequence of the two degenerate ground state in 
{\em trans}\,-polyacetylene, it is natural to expect the existence of 
excitations in polyacetylene in the form of topological solitons, or moving 
domain walls, separating the two degenerate minima. Depending on doping,
there exist 3 different types of topological solitons (kinks) in polyacetylene, 
having different quantum numbers. As shown in Fig. 14a, the removal of one 
electron from a polyacetylene chain creates a positively charged soliton 
with charge of +$e$ and spin zero [8]. The addition of one electron to a 
polyacetylene chain generates a neutral soliton with spin of 1/2, as shown 
in Fig. 14b. Figure 14c shows the case when two electrons are added to 
a polyacetylene chain. In this case, the soliton is negatively charged, having 
spin zero. Since these three kinds of solitons differ only by the occupation 
number of the zero energy state, the all have the same creation energy given 
by 
\begin{equation} 
E_{s} = \frac{2}{\pi} \Delta, 
\end{equation} 
where 2$\Delta$ is the width of the charge gap which separates the valence 
band and the conduction band. All three solitons occupy the midgap state of 
the charge gap, as schematically shown in Fig. 16b. Thus, to create a 
soliton is more energetically profitable than to add one electron to the 
conduction band, $E_{s} < 2 \Delta$. Figure 16a shows a $tanh$-shaped 
kink which connects the two electron bands and the electron density of 
states of the midgap state shown in Fig. 16b (compare with Fig. 6).
In Fig. 16a, the spatial density of states is 
determined by $| \psi (x)|^{2}$, where 
\begin{equation} 
\psi (x) = \frac{C}{\cosh(x/d)} = C \times sech(x/d),
\end{equation}  
and $C$ and $d$ are constants. 
\begin{figure}[t]
\epsfxsize=0.98\columnwidth
\centerline{\epsffile{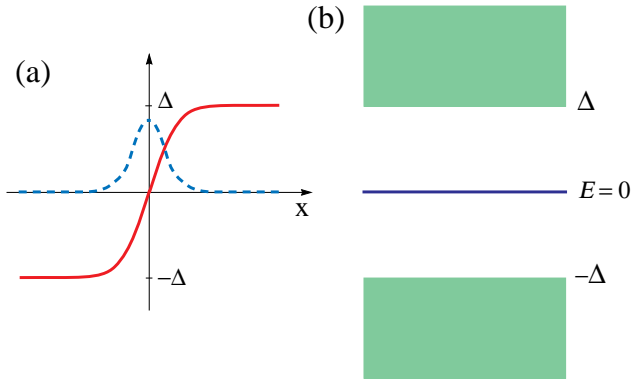}}
\caption{(a) A schematic plot of a topological soliton (thick curve) in 
{\em trans}\,-polyacetylene and the electron density of states 
$| \psi (x)|^{2}$ for the intragap state (dashed curve). The vertical axis for 
the density of states is not shown. (b) The spectrum of electron states for 
the soliton lattice configuration. The lower band corresponds to the valence 
band, and the upper band to the conduction band.}
\end{figure}

In Fig. 16a, the density of states is presented as a function of soliton 
position along the main polyacetylene axis. Figure 17 depicts the density 
of states in doped polyacetylene as a function of energy having the origin at 
the middle of the charge gap. The density of states shown in 
Fig. 17 corresponds to the total number of energy levels per unit volume 
which are available for possible occupation by electrons. Since the peak in the
spatial density of states, shown in Fig. 16a, is  not the delta function, but
has a finite width, then, according to the Fourier transform, the width of
corresponding peak in the spectral density of states is also finite. So, the
width of the soliton peak shown  schematically in Fig. 17 is finite. In Fig.
17, the height of the soliton  peak depends on the density of added or
removed electrons: the height  increases as the electron density increases [9].
However, one should realize  that there exists a maximum density of added or
removed electrons, above  which the charge gap collapses, and polyacetylene
becomes metallic.

The model was created in order to explain a very unusual behavior of the 
spin susceptibility in {\em trans}\,-polyacetylene. Upon doping, pristine 
polyacetylene becomes highly conducting. Strangely, the spin susceptibility 
of {\em trans}\,-polyacetylene remains small well into the conducting 
regime. Using the model, this strange behavior can be well understood: upon 
doping, the charge carriers in polyacetylene are not conventional electrons 
and holes, but spinless solitons (kinks). 

Following the original analysis [8], the model has been refined and it has 
been shown that the most probable defects are not the topological solitons 
but breathers (or polarons) which can be considered as a soliton-antisoliton 
bound state. In double-well potential, a kink can only be followed by an 
anti-kink, as shown in Fig. 15. Thus, the breather (polaron) solution is a 
sum of two {\em tanh} functions: 

\begin{equation} 
\Delta _{pol} = \Delta - \upsilon _{F} K \left[ \tanh \left( K \left( x + 
\frac{R}{2} \right) \right) - \tanh \left( K \left( x - \frac{R}{2} \right) \right)
\right], 
\end{equation}   
where $R$ is the distance between the soliton and antisoliton; $\upsilon
_{F}$  is the velocity on the Fermi surface, and $K$ is 
determined by 
\begin{equation} 
\upsilon _{F} K = \Delta \tanh (KR). 
\end{equation} 
The spectrum of electron states for $\Delta _{pol}$ consists of a valence 
band (with the highest energy -$\Delta$), a conduction band (with the lowest 
energy +$\Delta$), and two localized intragap states with energies
\begin{figure}[t]
\epsfxsize=0.8\columnwidth
\centerline{\epsffile{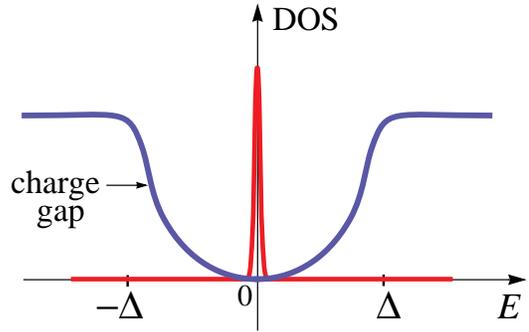}}
\caption{Sketch of the electron density of states (DOS) per unit energy 
interval in {\em trans}\,-polyacetylene for the soliton lattice configuration. 
The height of the soliton peak depends on the density of added or removed 
electrons [9].}
\end{figure} 
$\pm E_{0} (R)$, as shown in Fig. 18b, where 
\begin{equation}
E_{0} (R) = \frac{\Delta}{\cosh(KR)}.
\end{equation} 
The two intragap states are symmetric and antisymmetric superposition of 
the midgap state localized near the kink and antikink: 
\begin{equation} 
\psi _{\pm} (x) = \frac{1}{2} \left( \frac{\sqrt{K/2}}{\cosh(K(x - R/2))} 
|- \rangle \pm \frac{\sqrt{K/2}}{\cosh(K(x + R/2))}  |+ \rangle \right).
\end{equation} 
These results for the electron wave function hold for any distance $R$ 
between the kink and antikink. However, the configuration Eq. (28) is only 
self-consistent if 
\begin{equation} 
R_{sc} = \sqrt{2} \ln (1 + \sqrt{2}) \xi _{0}, 
\end{equation} 
where $\xi _{0} = \upsilon _{F} / \Delta$ denotes a correlation length. Then, 
$K \xi _{0} = 1/ \sqrt{2}$ and $E_{0} (R_{c}) = \Delta / \sqrt{2}$. 
\begin{figure}[t]
\epsfxsize=0.98\columnwidth
\centerline{\epsffile{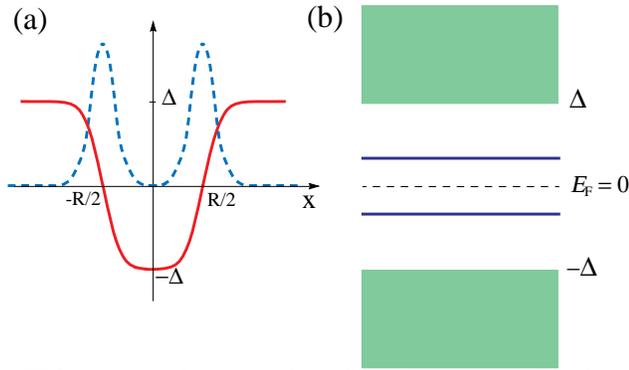}}
\caption{A schematic plot of a soliton-antisoliton lattice configuration 
(thick curve) in {\em trans}\,-polyacetylene and the electron density of 
states $| \psi (x)|^{2}$ for the two intragap state (dashed curve). The vertical 
axis for the density of states is not shown. (b) The spectrum of electron 
states for the soliton-antisoliton lattice configuration. The lower band 
corresponds to the valence band, and the upper band to the conduction band.}
\end{figure} 

The occupation of the two intragap states $\psi _{\pm}$ cannot be arbitrary: 
the solution is self-consistent if there is either one electron in the negative 
energy state $\psi _{-}$ and no electrons in the positive energy state 
$\psi _{+}$, or if $\psi _{-}$ is doubly occupied and $\psi _{+}$ is occupied 
by one electron. Only charged polarons are stable, thus they have charge $\pm 
e$ and spin 1/2. The polaron has the quantum numbers of an electron (a hole). 
In other words, the polaron is a bound state of a charged spinless soliton and 
a neutral soliton with spin 1/2. Figure 18a shows a bound state of a kink 
and an antikink separated by distance $R$, and the electron density of states 
of the two intragap states shown in Fig. 18b. 

By contrast, a charged kink and a charged antikink do not form a bound state, 
as they repel each other, even, in the absence of the Coulomb interaction. For 
a given density of added or removed electrons, this repulsion forces the 
charged kinks and antikinks to form a periodic lattice. At small density of 
added charges, the soliton lattice has a lower energy than the polaron lattice, 
since the kink creation energy $E_{s}$ [see Eq. (26)] is smaller than the energy 
of polaron creation, which is given by 
\begin{equation} 
E_{p} = \frac{2 \sqrt{2}}{\pi} \Delta = \sqrt{2} E_{s}.
\end{equation}  

In Fig. 18a, the density of states is presented as a function of 
soliton and antisoliton positions along the main polyacetylene axis. 
Figure 19 {\em schematically} depicts the density of states in doped 
polyacetylene as a function of energy having the origin at the middle of 
the charge gap. Since the occupation of the two intragap states cannot 
be arbitrary, the heights of the two peaks shown in Fig. 19 are, in fact, 
different: the right-hand peak should be lower than the left-hand peak. 
\begin{figure}[t]
\epsfxsize=0.8\columnwidth
\centerline{\epsffile{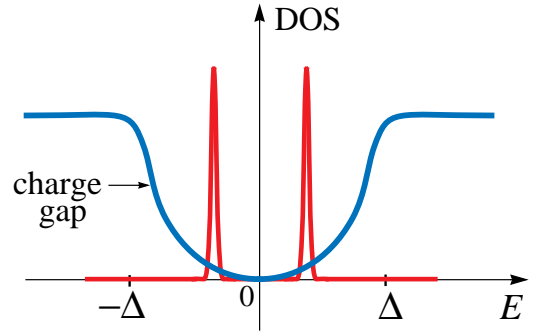}}
\caption{Sketch of the electron density of states (DOS) per unit energy 
interval in {\em trans}\,-polyacetylene for the soliton-antisoliton lattice 
configuration. The height of the soliton-antisoliton peaks depends on the 
density of added or removed electrons [9].}
\end{figure}  

Finally, it is worth emphasizing two aspects of kink-antikink bound state in 
any nonlinear system. First, the {\em distance} between the two peaks in the 
spectral density of states, shown in Figs. 18b and 19, depends on 
the {\em real distance} between the kink and the antikink, as schematically 
depicted in Fig. 18a. Second, the spatial {\em and} spectral densities of 
states of any kink-(anti)kink bound state (if such exists) are not sensitive to 
the ``polarity'' of the two kinks: the density of states of a
bound state of two  (anti)kinks (for example, in the periodic potential shown in 
Fig. 9) coincides  with the density of states of a kink-antikink bound state: 
compare the dashed  curves in Figs. 20 and 18a. 
\begin{figure}[h]
\epsfxsize=0.7\columnwidth
\centerline{\epsffile{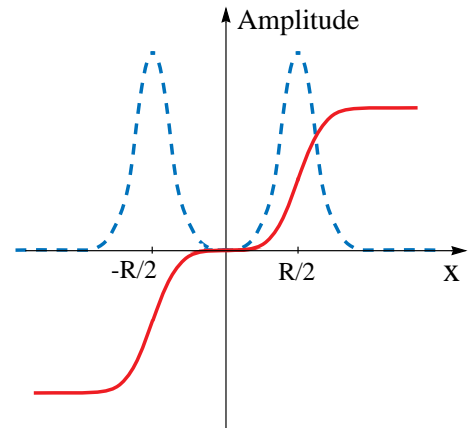}}
\caption{A schematic plot of two kinks (solid curve), for example, in 
infinite-well potential shown in Fig. 9, and the corresponding density of 
states (dashed curve). Compare with Fig. 18a.}
\end{figure} 

\section{Magnetic solitons} 

Solitons are not restricted to the macroscopic world---they exist on the 
microscopic scale too. 

Historically, in the theory of solid state magnetism, nonlinearities were 
considered small and were described in terms of linear theory. In the linear 
approximation, eigen-excitations in the magnetic medium were
considered as an ideal gas of non-interacting spin waves or
{\em magnons}. Taking into  account that nonlinearities in real medium
always present, the solution of  nonlinear equation of motion for magnetic excitations
represents a  magnetic soliton. The term ``magnetic soliton'' has a broad meaning:
there  exist magnetic kinks and envelope solitons which are described by the 
sine-Gordon and NLS equations, respectively. 

Magnetic {\em kink} solitons clearly show up in the properties of 
quasi-one-dimensional magnetic materials in which spins interact strongly 
along one axis of the crystal and very weakly along the other axes. Such spin 
chains are qualitatively similar to the pendulum chain shown in Fig. 7. Let 
us assume that the pendulums in Fig. 7 represents spins in a spin chain. 
Such a situation when all spins in the chain are oriented in the same 
direction (down in Fig. 7) corresponds to a {\em ferromagnetic} ordering. 
And a torsion of the spin lattice, shown in Fig. 7, can propagate as a soliton 
in the crystal. An {\em antiferromagnetic} ordering represents a staggered 
spin orientation in the chain, i.e. for each spin in the chain, its nearest 
neighbors have the opposite orientation. In antiferromagnetically ordered 
materials, a torsion of the spin lattice also corresponds to a magnetic 
kink-soliton. Magnetic kink-solitons are domain walls which can be moving 
or stationary, and described by the sine-Gordon equation. These solitons are 
usually created thermally and their dynamics can be studied very accurately 
by using neutron scattering. The soliton concept is again precious
because the  solitons can be treated as quasiparticles and the results can be obtained
by  investigating a ``soliton gas.'' 

For a qualitative understanding of magnetic solitons, we need not look into  
details of so-called {\em exchange} forces between spins. As distinct from 
quantum exchange forces that really act among the nearest neighbors, these 
exchange forces are classical and have a {\em long range}. Accordingly, it 
becomes energetically favorable for ferromagnetically ordered spins to 
divide into many groups. In one half of these groups, the spins look up, in the 
other half, they look down. These groups are also called {\em domains}, and 
the borders between domains---domain walls which are similar to the
Frenkel-Kontorova dislocations. Therefore, they are solitons, and their 
evolution is described by the sine-Gordon equation. Like dislocations, the 
domain walls may move along the crystal if there is no obstruction from the 
crystal imperfections or from other domain walls. 

The latter domain walls are somewhat easier to observe than the 
dislocations. A typical distance between two magnetic domain walls is about 
1 mm, and their thickness is typically 10 $\mu$m or so. Covering 
well-polished surface of a ferromagnetic sample with a thin powder of a 
magnetic material, it turns out that its particles gather near the domain 
walls. Thus, {\em the particles are attracted by the domain walls}.

As mentioned above, there also exist magnetic {\em envelope} solitons. In 
the linear approximation, a propagating linear packet of spin waves is spread 
by dispersion because individual spin waves in the packet are totally 
independent of each other. The solution of equation of motion for the 
magnetization, which contains nonlinear effects, represents
a  magnetic envelope soliton. The latter is a cluster or a bound state
of large  number of spin waves where the attraction between spin waves compensate 
the spreading effect of dispersion. The existence of spin wave
envelope  solitons was predicted theoretically, and later they were experimentally 
observed in the microwave frequency range. This type of magnetic solitons 
are described by the NLS equation. 

\section{Self-trapped states: the Davydov soliton} 

Solitons represent localized states and, in real systems, there exist 
{\em self-localized} states. In other words, there are cases when, in 
exchange of interaction with a medium, an excitation or a particle locally 
deforms the medium in a way that it is attracted by the deformation. In the 
context of condensed matter physics, Landau proposed that an electron in a 
solid can be considered as being ``dressed'' by its self-created polarization 
field, forming a quasiparticle. When the particle-field interaction 
(electron-phonon coupling) is strong, both the particle wave function and 
lattice deformation are localized. In the three-dimensional case, this 
localized entity is known as a {\small} (Holstein) polaron and, in one 
dimension a Davydov soliton or {\em electrosoliton}. Their integrity is 
maintained owing to the dynamical balance between the dispersion (exchange 
inter-site interaction) and the nonlinearity (phonon-electron coupling). Figure 
21 schematically shows a self-trapped state of a particle (a small polaron 
or a Davydov soliton). In a self-trapped state, {\em both} the particle-wave 
function and lattice deformation are {\em localized}.
\vspace{3mm} 
\begin{figure}[t]
\epsfxsize=0.9\columnwidth
\centerline{\epsffile{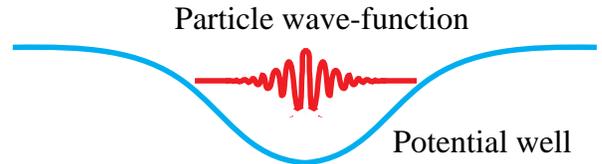}}
\caption{Sketch of a self-trapped state: a particle 
(electron or hole) with a given wave function deforms the lattice inducting 
a potential well which in turn traps the particle.}
\end{figure}  

Before we discuss the Davydov bisoliton model of high-$T_{c}$ 
superconductivity, we consider first the concept of the  Davydov soliton 
[10, 11]. 

To describe the self-focusing phenomena, the NLS equation is used, 
\begin{equation} 
\left[ i\hbar \frac{\partial}{\partial t} + \frac{\hbar ^{2}}{2m} 
\frac{\partial ^{2}}{\partial x^{2}} + G| \psi (x, t)|^{2} \right] \psi (x, t) = 0. 
\end{equation} 
The equation is written in the long-wave approximation when the excitation 
wavelength $\lambda$ is much larger than the characteristic dimension of 
discreteness in the system, i.e. under the condition $ka = 2 \pi a/ \lambda \ll$ 
1. The equation describes the complex field $\psi (x, t)$ with self-interaction. 
The function $| \psi |^{2}$ determines the position of a quasiparticle of mass 
$m$. The second term in the NLS equation is responsible for the
dispersion,  and the third one for nonlinearity. The coefficient $G$ characterizes 
the intensity of the nonlinearity. 

When the nonlinearity is absent ($G$ = 0), the NLS equation has solutions in 
the form of plane waves, 
\begin{equation} 
\psi (x, t) = \Phi _{0} \exp [i(ka - \omega (k)t)],
\end{equation} 
with the square dispersion law $\omega (k) = \hbar k^{2}/2m$. 
With nonlinearity ($G \neq$ 0) in the system having a translational 
invariance, the excited states move with constant velocity $V$. Therefore 
it is convenient to study solutions of the NLS equation in the reference 
frame 
\begin{equation} 
\zeta = (x - Vt)/a, 
\end{equation} 
moving with constant velocity. In this reference frame the NLS equation has 
solutions in the form of a complex function
\begin{equation} 
\psi (x, t) = \Phi ( \zeta ) \exp [i(kx - \omega t)], \ k = mV/ \hbar, 
\end{equation} 
where the real function $\Phi ( \zeta )$ satisfies the nonlinear equation 
\begin{equation} 
\left[ \hbar \omega - \frac{1}{2} mV^{2} + J \frac{\partial ^{2}}{\partial 
x^{2}} + G_{0} \Phi ^{2} ( \zeta ) \right] \Phi ( \zeta ) = 0, 
\end{equation} 
where $J = \hbar ^{2}/2ma^{2}$. 

The last equation has two types of solutions: nonlocalized and localized ones. 
The nonlocalized solution corresponds to a constant value of the amplitude 
$\Phi ( \zeta ) = \Phi _{0}$ and the dispersion law 
\begin{equation} 
\hbar \omega = \frac{1}{2} mV^{2} - G_{0} \Phi _{0}^{2}.
\end{equation} 
If a particle is at a distance $L$, then $\Phi _{0}^{2} = L^{-1}$ and, in the 
limit $L \rightarrow \infty$, the second term in the last equation tends to 
zero. 

The localized solution of the NLS equation normalized by the condition 
\begin{equation} 
\int \Phi ^{2} ( \zeta ) d \zeta = 1, 
\end{equation} 
is represented, as we already know, by the bell-shaped function 
\begin{equation} 
\Phi ( \zeta ) = \sqrt{\frac{g}{2}} \times sech(g \zeta), 
\end{equation} 
with the dimensionless nonlinearity parameter $g$ given by
\begin{equation} 
g = G_{0}/4J. 
\end{equation} 
The function $\Phi ( \zeta )$ is nonzero on a segment $\Delta \zeta \approx 
2 \pi /g$. The larger the nonlinearity parameter the smaller is the 
localization region. 

The energy of localized excitations is determined by the expression 
\begin{equation} 
\hbar \omega = \frac{1}{2} mV^{2} - Jg^{2}.
\end{equation} 

In the localized solution found above, the nonlinearity is generated by the 
``self-action'' of the field. Let us now study a self-trapping effect when 
two linear system interact with each other. As an example, we consider the 
excess electron in a quasi-one-dimensional atomic (molecular) chain. If 
neutral atoms (molecules) are rigidly fixed in periodically arranged sites 
$na$ of a one-dimensional chain, then, due to the translational invariance of 
the system, the lowest energy states of an excess electron are determined 
by the conduction band. The latter is caused by the electron collectivization. 
In the continuum approximation, the influence of a periodic potential is taken 
into account by replacing the electron mass $m_{e}$ by the effective mass 
$m = \hbar /2a^{2}J$ which is inversely proportional to the exchange 
interaction energy that characterizes the electron jump from one node site 
into another. In this approximation, the electron motion along an ideal chain 
corresponds to the free motion of a quasiparticle with effective mass $m$ 
and electron charge. 

Taking into account of small displacements of molecules of mass $M$ 
($\gg m$) from their periodic equilibrium positions, there arises the 
short-range deformation interaction of quasiparticles with these 
displacements. When the deformation interaction is rather strong, the 
quasiparticle is self-localized. Local displacement caused by a quasiparticle 
is manifest as a potential well that contains the particle, as 
schematically shown in Fig. 21. In turn, the quasiparticle deepens the well. 

A self-trapped state can be described by two coupled differential equations 
for the field $\psi (x, t)$ that determines the position of a quasiparticle, and 
the field $\rho (x, t)$ that characterizes local deformation of the chain and 
determines the decrease in the relative distance $a \rightarrow a - \rho(x,t)$ 
between molecules of the chain, 
\begin{equation} 
\left[ i \hbar \frac{\partial}{\partial t} + \frac{\hbar ^{2}}{2m} 
\frac{\partial ^{2}}{\partial x^{2}} + \sigma \rho (x, t) \right] \psi (x, t) = 0,  
\end{equation} 
\begin{equation} 
\left( \frac{\partial ^{2}}{\partial t^{2}} - c_{0}^{2} 
\frac{\partial ^{2}}{\partial x^{2}} \right) \rho (x, t) - 
\frac{a^{2} \sigma}{M} \frac{\partial ^{2}}{\partial x^{2}} | \psi (x, t)|^{2} = 0.  
\end{equation} 

The first equation characterizes the motion of a quasiparticle in the local 
deformation potential $U = - \sigma \rho (x, t)$. The second equation 
determines the field of the local deformation caused by a quasiparticle. 
The system of the two equations are connected through the parameter 
$\sigma$ of the interaction between a quasiparticle and local deformation. 
The quantity $c_{0} = a \sqrt{k/M}$ is the longitudinal sound velocity in the 
chain with elasticity coefficient $k$. When there is 
one quasiparticle in the chain, the function $\psi (x, t)$ is 
normalized by 
\begin{equation} 
\frac{1}{a} \int \limits_{- \infty}^{\infty} | \psi (x, t)|^{2} dx = 1. 
\end{equation} 

In the reference frame $\zeta = (x - Vt)/a$ moving with constant velocity 
$V$, the following equality 
$\partial \rho (x, t)/ \partial t = -V/a \times \partial \rho / \partial \zeta$ 
holds. Then, the solution for $\rho (x, t)$ has the form 
\begin{equation} 
\rho (x, t) = \frac{ \sigma }{k(1 - s^{2})} | \psi (x, t)|^{2}, 
\ s^{2} = V^{2}/c_{0}^{2} \ll 1.  
\end{equation} 

Substituting the values $\rho (x, t)$ into Eq. (44), we transform it to a 
nonlinear equation for the function $\psi (x, t)$,
\begin{equation} 
\left[ i \hbar \frac{\partial}{\partial t} + \frac{\hbar ^{2}}{2m} 
\frac{\partial ^{2}}{\partial x^{2}} + 2gJ| \psi (x, t) |^{2} \right] \psi (x, t) = 0. 
\end{equation}  
where 
\begin{equation} 
g \equiv \frac{\sigma ^{2}}{2k(1 - s^{2})J} 
\end{equation} 
is the dimensional parameter of the interaction of a quasiparticle with a local 
deformation. Substituting the function 
\begin{equation} 
\psi (x, t) = \Phi ( \zeta ) \exp [i(kx - \omega t)], \ k = mV/ \hbar 
\end{equation} 
into the Eq. (48) we get the equation 
\begin{equation} 
[ \hbar \omega - \frac{1}{2} mV^{2} - J \frac{\partial ^{2}}{\partial \zeta ^{2}} 
+ 2gJ \Phi ^{2} ( \zeta )] \Phi ( \zeta ) = 0,   
\end{equation} 
for the amplitude function $\Phi ( \zeta )$ normalized by 
$\int \Phi ^{2} ( \zeta )d \zeta = 1$. The solution of this equation is 
\begin{equation} 
\Phi ( \zeta ) = \frac{1}{2} \sqrt{g} \times sech(g \zeta /2), 
\end{equation} 
with the value 
\begin{equation} 
\hbar \omega = \frac{1}{2} mV^{2} - D(s). 
\end{equation} 
The quantity 
\begin{equation} 
D(s) = \frac{1}{8} g^{2} J
\end{equation} 
determines the binding energy of a particle and the chain deformation 
produced by the particle itself. According to Eq. (52), a quasiparticle is 
localized in moving reference frame 
\begin{equation} 
\Delta \zeta = 2 \pi /g, 
\end{equation} 
as shown in Fig. 22. In this region, the field localization is 
characterized by the function 
\begin{equation} 
\rho ( \zeta ) =  \frac{g \sigma}{4k(1 - s^{2})} sech^{2}(g \zeta /2). 
\end{equation} 
The following energy is necessary for deformation: 
\begin{equation} 
W = \frac{1}{2} k(1 + s^{2}) \int \rho ^{2} ( \zeta ) d \zeta = 
\frac{1}{24} g^{2} J(1 + s^{2}).  
\end{equation} 

Measured from the bottom of the conduction band of a free quasiparticle, the 
total energy (including that of deformation) transferred by a soliton moving 
with velocity $V$ is determined by the expression 
\begin{equation} 
E_{s} (V) = W + \hbar \omega = E_{s} (0) + \frac{1}{2} M_{sol} V^{2}, 
\end{equation} 
in which the energy of a soliton at rest $E_{s} (0)$ and its effective mass 
\begin{figure}[t]
\epsfxsize=0.9\columnwidth
\centerline{\epsffile{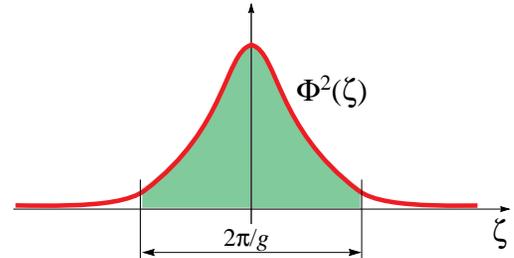}}
\caption{A schematic plot of a self-trapped soliton. The soliton is very 
localized, having a size of 2$\pi /g$.}
\end{figure} 
$M_{sol}$ are determined, respectively, by the equalities 
\begin{equation} 
E_{s} (0) = \frac{1}{12} g^{2}J, 
\end{equation} 
\begin{equation} 
M_{sol} = m(1 + \frac{g^{2}J}{3a^{2} k}). 
\end{equation} 
The soliton mass $M_{sol}$ exceeds the effective mas of a quasiparticle $m$, 
as its motion is accompanied by the motion of a local deformation. 

The effective potential well where quasiparticles are placed, is determined, 
in the reference frame $\zeta$, by the expression 
\begin{equation} 
U = - \sigma \rho (x, t) = - g^{2}J \times sech^{2} (g \zeta /2).
\end{equation} 

The self-trapped soliton is very stable. The soliton moves with the 
velocity $V < c_{0}$, otherwise, the local deformation of the chain will 
be not able to follow the quasiparticle. Alternatively, the soliton can be 
stationary. 

It is worth emphasizing that the Davydov solitons are conceptually 
different from the small polarons which are three-dimensional. In fact, 
the small polarons are practically at rest owing to their large mass. 

The bisoliton excitations are discussed separately (Chapter 7 in [14]).  

\section{Discrete breathers} 

In lattices, intrinsic localized modes with internal (envelope) oscillations 
(see Fig. 11c) are often called {\em discrete breathers} because an envelope 
mode or nonlinear wavepacket can be considered as the small amplitude 
limit of a breather-solution for the continuum sine-Gordon equation and 
also for other systems in the semi-discrete approximation, where the 
oscillations (discrete carrier wave) vary rapidly inside the slow 
(continuous) envelope. In this approximation, discrete breathers are linked 
to envelope solitons. Contrary to continuous breathers which are known to 
exist only in some particular systems, discrete breathers are structurally 
stable as soon as nonlinear oscillators are coupled 
sufficiently weakly and locally. 

In purely harmonic lattices, spatially localized modes can occur only when 
defects or disorder are present, so that the translational invariance of the 
underlying lattice is broken. Discrete breathers may be created anywhere in 
a perfect homogenous nonlinear lattice. As an example, let us consider 
the behavior of a one-dimensional chain of particles with mass $m$, bound 
by massless springs. A similar chain was used in the computer simulations 
by Fermi, Pasta and Ulam (see above). In this chain, a discrete breather is a 
nonlinear excitation centered at a lattice site, which involves longitudinal 
displacements of a few masses. The longitudinal displacements of any two 
neighboring masses are in antiphase, so that the displacement of the 
central mass is maximal, and two neighboring masses have smaller, 
antiphase displacements relatively to the central mass etc. This case 
corresponds to an {\em odd parity mode}, however, one may also have an 
{\em even parity mode} centered at the midpoint between two masses. 

The basic properties of the discrete breather or the {\em intrinsic localized 
mode} can be summarized as follows [7]: 

\begin{itemize} 

\item[i.] 
It is a time periodic solution localized in space which may occur in one, 
two or three dimensions. 

\item[ii.]
It may be centered on any lattice site giving rise to a configurational 
entropy. 

\item[iii.]
It extends over a few lattice sites. 

\item[iv.]
It has an amplitude dependent frequency. 

\item[v.]
It may be mobile. 

\end{itemize} 

Contrary to solitons, discrete breathers do not require integrability for 
their existence and stability. A particular important condition of breather 
existence is that all multiples of the fundamental frequency lie outside of 
the linear excitations spectrum. Discrete breathers are predicted to exist; 
however, they have not yet been observed experimentally. 

\section{Structural phase transitions} 

From Chapter 3 we know that the phase diagram of cuprates is very rich on 
structural phase transitions. What is a structural phase transition? A 
structural phase transition corresponds to the shifts of the equilibrium 
positions of the atoms (molecules). They take place owing to the instability 
of some lattice displacement pattern, which drives the system from some 
stable high-temperature phase at $T > T_{d}$ to a different 
low-temperature lattice configuration at $T < T_{d}$, where $T_{d}$ is the 
critical temperature. The dynamics of such transitions is frequently 
characterized by a vibrational or phonon mode whose frequency sharply 
decreases as the temperature approaches the critical temperature $T_{d}$ 
from above. As a result, the restoring force corresponding to that 
displacement pattern softens, and one calls this particular mode a 
{\em soft mode}. 

At the critical temperature $T_{d}$, the lattice displacements become 
large and {\em the dynamics is highly nonlinear}. As a consequence, 
equations describing the structural phase transition become nonlinear. 
Simulating structural phase transitions in one-dimensional lattice, a model 
corresponding to a structural phase transition can be presented by a large 
number of double-well potentials shown in Fig. 15, forming a 
one-dimensional chain similar to that in Fig. 5. Contrary to the case in Fig. 
5, in the latter model, the balls (atoms) cannot jump from one double-well 
to another, however, being able to change the minima in each double-well 
potential. Writing down the Hamiltonian, the solution of the obtained 
nonlinear equation strongly depends on the coupling (spring) strength 
between nearest-neighbor balls (atoms) and the potential-barrier height 
inside each double well. In the case of strong coupling between balls 
(atoms), the equation takes a form of the sine-Gordon equation. The 
solution of the equation coincides with Eq. (9), having a $tanh$ shape and 
corresponding to a localized kink soliton. 

So, the main point of this section is that a structural 
phase transition can be presented as a localized soliton (domain wall).  
Owing to preparation, impurity and defects content, the crystal may be in 
a state of coexistence of two or more structural phases, forming 
{\em domains} separated by {\em walls}. 

\section{Tunneling and the soliton theory} 

What can we expect from tunneling measurements in a system having 
solitons? 

As discussed in Chapter 2 in [14], a tunneling conductance $dI(V)/dV$ obtained 
in a superconductor-insulator-normal metal (SIN) junction directly relates to 
the electron density of states per unit energy interval since voltage $V$ 
multiplied  by $e$ represents energy. All the soliton solutions 
considered above are  real-space functions, i.e. functions of $x$ and $t$. 
Therefore, before we discuss tunneling measurements in terms of the 
solitons solutions, we need to transform the solutions from real space 
into momentum realm. 

First, let us {\em qualitatively} determine the shape of tunneling 
characteristics caused by solitons. Taking into account that $V = E/e$, 
Figure 17 is represented in conductance units, $dI/dV$ vs $V$, in 
Fig. 23a. Bearing in mind that tunneling current is the sum under the 
conductance curve, $I(V) = \int \frac{dI(V)}{dV}dV + C$, where $C$ is the 
constant, the $I(V)$ characteristic of the soliton peak 
shown in Fig. 23a at zero bias is schematically depicted in Fig. 23b. 
The constant $C$ is defined by the condition $I(V= 0) = 0$. 
\begin{figure}[t]
\epsfxsize=0.98\columnwidth
\centerline{\epsffile{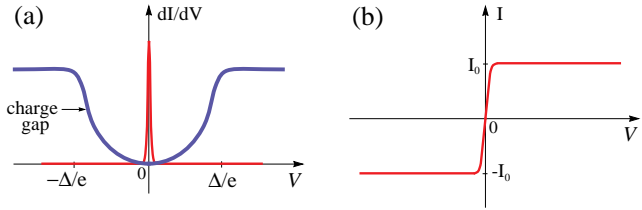}}
\caption{(a) Same as Fig. 17 but in conductance units: $dI/dV$ 
vs $V$ (for an SIN junction); and (b) the $I(V)$ characteristic 
corresponding to the soliton peak in plot (a). The $I(V)$ 
characteristic of the charge gap is not shown.}
\end{figure} 

The same procedure has been done for two polaron peaks shown in Fig. 19: the 
$dI(V)/dV$ and $I(V)$ characteristics of a polaron are shown in Figs. 24a 
and 5.24b, respectively. 

Let us now {\em quantitatively} determine the shape of tunneling 
conductance peak, obtained in a system having topological solitons. In 
general, the total energy of a soliton consists of the energy of a static 
soliton and  its kinetic energy. For simplicity, consider a static soliton, 
thus, a soliton  with a velocity $\upsilon$ = 0. The spatial density of states 
of a topological  soliton is determined by $| \psi (x)|^{2}$, where $\psi (x) = 
C \times sech(x/d)$  given by Eq. (27). Then, the spectral function 
$\varphi (k)$ can be found by  applying the Fourier transform to $\psi (x)$, 
where $k$ is the wave number: 
\begin{equation} 
\varphi (k) = \int \limits_{- \infty}^{\infty} \psi (x) e^{-ikx}dx.  
\end{equation}  
Bearing in mind that the Fourier transform of the $sech( \pi x)$ function 
gives $sech( \pi k)$ [12], then the energy spectrum of 
a topological soliton is also enveloped by the $sech$ hyperbolic function, 
centered at  some $k_{0}$. Then, for $\Delta k = (k - k_{0})$, we have 
$\varphi ( \Delta k) = C_{1} \times sech(d_{1} \Delta k)$,  where $C_{1}$ 
and $d_{1}$ are constants. 

Since $E = \frac{\hbar ^{2}k^{2}}{2m}$, where $m$
is the  mass, then 
\begin{equation} 
E(k) - E(k_{0}) = \Delta E = \frac{\hbar ^{2}k^{2}}{2m} - 
\frac{\hbar ^{2}k_{0}^{2}}{2m} \simeq \frac{\hbar ^{2}k_{0}}{m} \Delta k 
\end{equation} 
valid in the vicinity of $k_{0}$, i.e. for $\Delta k \ll k_{0}$. Substituting 
$\Delta k \rightarrow \Delta E$, one obtains 
\begin{equation} 
\varphi (\Delta E) = C_{1} \times sech(d_{1} \frac{m}{\hbar ^{2}k_{0}} 
\Delta E).  
\end{equation} 
\begin{figure}[t]
\epsfxsize=0.98\columnwidth
\centerline{\epsffile{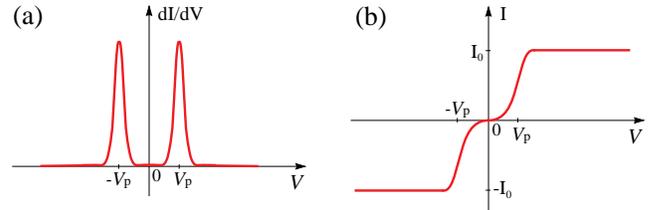}}
\caption{(a) Same as Fig. 19 but in conductance units: $dI/dV$ 
vs $V$ (for an SIN junction). The charge gap is not shown, and the width of 
the conductance peaks is slightly enlarged in comparison with that in 
Fig. 19. (b) The corresponding $I(V)$ characteristic.}
\end{figure} 
Assuming that the soliton energy remains constant, i.e. $k_{0}$ = constant, 
then, the spectral density of states is determined by
\begin{equation} 
| \varphi (\Delta E)|^{2} = C_{1}^{2} \times sech^{2} (\Delta E/E_{0}), 
\end{equation} 
where $E_{0} = \frac{\hbar ^{2}k_{0}}{md_{1}}$. Thus, in the vicinity of 
peak energy, the energy density of states is proportional to 
$sech^{2} (\Delta E/E_{0})$. 

Finally,  in SIN tunneling measurements performed in a system having topological 
solitons, for example, in polyacetylene, one can expect that, near peak bias,  
\begin{equation} 
\frac{dI(V)}{dV} \simeq A \times sech^{2} (V/V_{0}) 
\end{equation} 
and 
\begin{equation} 
I(V) \simeq B \times \tanh (V/V_{0}), 
\end{equation} 
where $V$ is the applied bias, and $A$, $B$ and $V_{0}$ are the constant. 
In the equations, $\Delta V \rightarrow V$ since the peak is centered at 
zero bias, as shown in Fig. 23a. 

By the same token, for the polaron characteristics shown in Fig. 24, one 
can obtain that, in an SIN junction,  
\begin{equation} 
\frac{dI(V)}{dV} \simeq A \times \left[ sech^{2} \left( 
\frac{V + V_{p}}{V_{0}} \right) + sech^{2} \left( \frac{V - V_{p}}{V_{0}} 
\right) \right] 
\end{equation} 
and 
\begin{equation} 
I(V) \simeq B \times \left[ \tanh \left( \frac{V + V_{p}}{V_{0}} \right) + 
\tanh \left( \frac{V - V_{p}}{V_{0}} \right) \right], 
\end{equation} 
where $V_{p}$ is the peak bias, as shown in Fig. 24. The latter equations 
are valid in the vicinities of $\pm V_{p}$. 

What is interesting is that the results obtained above for a topological 
soliton are also valid for a static envelope soliton which is 
schematically shown in Fig. 25a. As discussed above, in real space 
coordinates, the enveloping curve of an envelope soliton has the 
$sech$\,-function shape 
(the dashed curve in Fig. 25a). The energy spectrum of the envelope 
soliton is schematically presented in Fig. 25b. From the Fourier transform, 
one can easily obtain that the enveloping curve of energy spectrum of an 
envelope soliton, shown in Fig. 25b by the dashed line, is also the $sech$ 
hyperbolic function. Hence, for a static envelope soliton, the 
density-of-state peak of energy spectrum has the $sech^{2}$\,-function 
shape, as for a topological soliton. 
\begin{figure}[t]
\epsfxsize=0.98\columnwidth
\centerline{\epsffile{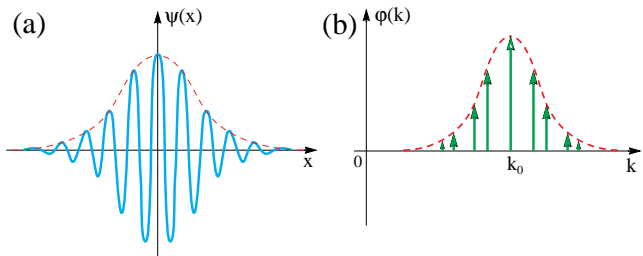}}
\caption{(a) An envelope soliton in real space, and (b) its energy 
spectrum, shown schematically.}
\end{figure} 

Thus, quasiparticle peaks in a conductance obtained in an SIN tunneling 
junction in a system having either envelope or topological solitons can be 
fitted by the $sech^{2}$ hyperbolic function, and the corresponding 
tunneling $I(V)$ characteristic by the $tanh$ hyperbolic function. 
As a consequence, quasiparticle peaks in a conductance obtained in a 
superconductor-insulator-superconductor (SIS) tunneling junction have to 
be fitted by the convolution of the $sech^{2}$ function with itself [see 
Eq. (18)]. The convolution cannot be resolved analytically. As shown in Chapter 12 
in [14], the $sech^{2}$\,- and $tanh$\,-function fits can be applied also 
to tunneling spectra obtained in an SIS junction. Therefore, independently from 
the type of tunneling junction, SIN or SIS, the $sech^{2}$ function can be used 
to fit quasiparticle peaks in conductances, and the $tanh$ function to fit 
$I(V)$ characteristics, caused by solitons. However, one has to realize 
that quasiparticle conductance peaks caused by solitonic states will appear 
in the ``background'' originated from other electronic states present in the system. 

For a moving soliton, i.e. for $\upsilon \neq$ 0, the results
obtained  above are also valid if its velocity $\upsilon$ is small. 

\section{Modern solitons} 

The modern soliton theory is a very broad area of research not only in 
different branches of physics but in different branches of science, such as 
chemistry, biology, medicine, astronomy etc. Practical devices based on the 
soliton concept, for example, in optics are now a multimillion-dollar 
industry. 

In 1962, that is even before Kruskal and Zabusky's work, Skyrme suggested 
that elementary particles, for example, protons can be regarded as solitons 
(in modern usage). He used kink solitons being 
the exact solutions of the sine-Gordon equation to describe the dynamics of 
elementary particles. He demonstrated that his solitons behave like 
fermions. At that time, his ideas were very unpopular. However, rigorous 
proofs of the equivalence between the sine-Gordon theory and the theory of 
fermions were given only 15 years later by others. 

Attempts to use optical pulses for transmitting information started as soon 
as good enough quality optical fibers became available. Since even 
superb quality fibers have dispersion, it is a more or less evident idea to use 
optical solitons for the transmission of information. The nonlinearity which 
is normally very small in optical phenomena results in the self focusing of 
the laser beam. As we already know, the self-focusing phenomena are 
described by the NLS equation. Thus, the optical soliton is in fact an envelope 
soliton. In optics, the envelope soliton shown in Fig. 11c is called a {\em 
bright soliton} because it corresponds to a pulse of light. There also exists 
another type of envelope solitons which are called in optics {\em dark 
solitons}, because they correspond to a hole in the continuous light (carrier) 
wave.  Using solitons for transmitting the information, the transmission 
speed of optical fibers is enormous, and can be about 100 Gigabits per 
second. In addition, soliton communication is more 
reliable and less expensive.

Solitons are encountered in biological systems in which the nonlinear
effects are often the predominant ones [13]. 
For example, many biological reactions would not occur without large 
conformational changes which cannot be described, even approximately, as 
a superposition of the normal modes of the linear theory. 

The shape of a nerve pulse was determined more than 100 years ago. 
The nerve pulse has a bell-like shape and propagates with the velocity
of about 100 km/h. The diameter of nerves in mammals is less than 20 
microns and, in first approximation, can be considered as 
one-dimensional. For almost a century, nobody realized that the nerve pulse 
is the soliton. So, all living creatures including humans are literally stuffed 
by solitons. Living organisms are mainly organic and, in principle, should be 
insulants---solitons are what keeps us alive.  

The last statement is true in every sense: the blood-pressure pulse seems to 
be some kind of solitary wave [7]. The muscle contractions are stimulated 
by solitons [11]. 

The so-called {\em Raman effect} is closely related to supplying solitons 
with additional energy. The essence of the Raman effect is that the spectrum 
of the scattered (diffused) light is changed by its interaction with the 
molecules of the medium. Crudely speaking, the incoming wave is modulated 
by vibrating molecules. These vibrations are excited by the wave, but the 
frequency of the vibrations depends only on properties of 
the molecules. 

So, I could continue to enumerate different cases and different systems where 
solitons exist. Nevertheless, I think that it is already enough information for 
the reader to understand the concept of the soliton, and to perceive them as 
real objects because we deal with them every day of our lives. 

\section{Neither a wave nor a particle} 

At the end, I have a proposal. Solitons are often considered as ``new objects 
of Nature'' [5]. It is not true in the global sense. Solitons exist since the 
existence of the universe. It is true that solitons are ``new objects of Nature'' 
{\em for humans}. However, they existed even before the life began. 

Solitons are waves, however, very strange waves. They are not particles, 
however, they have particle-like properties. They transfer energy, and do it 
very efficiently. Since the formulation of the relativistic theory, we know 
that the mass is an equivalent of energy, and relates to it by 
\[ E = mc^{2}. \]
From this expression, it is not important if an object has a mass or 
not: it is an object if it has an ability to transfer (localized) energy. 

In my opinion, solitons do not have to be associated neither with waves nor 
with particles. They have to remain as {\em solitons}. It is simply a 
{\em separate} (and very old) form of the existence of matter. And, they 
should remain as they are. 

I am sure that, in the future, we shall add to the group of {\em waves, 
solitons}, and {\em particles} new names---new forms of the existence of 
matter. However, it will happens in the future and, right now, let us return 
to our problem---the mechanism of high-$T_{c}$ superconductivity in 
cuprates. 

\vspace*{-1mm}

\section{APPENDIX: Books recommended for further reading}

\vspace*{-2mm}

\noindent
- M. Remoissenet, {\it Waves Called Solitons} (Springer-Verlag, Berlin, 
1999). 

\noindent
- A. S. Davydov, {\it Solitons in Molecular Systems} (Kluwer Academic, 
Dordrecht, 1991). 

\noindent
- A. T. Filippov, {\it The Versatile Soliton} (Birkh\"auser, Boston, 2000). 

\noindent
- G. Eilenberger, {\it Solitons} (Springer, New-York, 1981). 

\noindent
- F. K. Kneub\"uhl, {\it Oscillations and Waves} (Springer, Berlin, 1997). 

\noindent
- {\it Solitons and Condensed Matter Physics}, A. R. Bishop and T. Schneider, 
(ed.) (Springer, Berlin, 1978). 

\noindent
- {\it Nonlinearity in Condensed Matter}, A. R. Bishop, D. K. Cambell, P. Kumar, 
and S. E. Trullinger, (ed.) (Springer, Berlin, 1987). 

\noindent
- M. Lakshmanan, {\it Solitons} (Springer, Berlin, 1988). 


\end{document}